\documentclass[a4paper,11pt]{article}
\pdfoutput=1 % if your are submitting a pdflatex (i.e. if you have
\usepackage{jheppub} % for details on the use of the package, please
\usepackage{amsmath,amssymb,amsthm,amscd,graphicx}
\usepackage{slashed}
\usepackage[compat=1.1.0]{tikz-feynman}
\input epsf.sty

\theoremstyle{definition}

\newcommand{\CB}{{\cal B}}
\newcommand{\CC}{{\cal C}}

\newcommand{\CF}{{\cal F}}

\newcommand{\CH}{{\cal H}}
\newcommand{\CI}{{\cal I}}

\newcommand{\CL}{{\cal L}}
\newcommand{\CM}{{\cal M}}

\newcommand{\CO}{{\cal O}}
\newcommand{\CP}{{\cal P}}

\newcommand{\CR}{{\cal R}}

\newcommand{\CT}{{\cal T}}

\def\IZ{{\mathbb Z}}

\newcommand{\vev}[1]{\left\langle #1 \right\rangle}

\newcommand{\bpsi}{\boldsymbol{\psi}}

\newcommand{\tr}{{\rm Tr}}
\newcommand{\re}{{\rm e}}
\newcommand{\ri}{{\rm i}}
\newcommand{\rd}{{\rm d}}

\newcommand{\be}{\begin{equation}}
\newcommand{\ee}{\end{equation}}
\newcommand{\ba}{\begin{aligned}}
\newcommand{\ea}{\end{aligned}}
\newcommand{\ben}{\begin{eqnarray}\displaystyle}
\newcommand{\een}{\end{eqnarray}}

\newcommand{\sectiono}[1]{\section{#1}\setcounter{equation}{0}}

\newdimen\tableauside\tableauside=1.0ex
\newdimen\tableaurule\tableaurule=0.4pt
\newdimen\tableaustep
\def\phantomhrule#1{\hbox{\vbox to0pt{\hrule height\tableaurule width#1\vss}}}
\def\phantomvrule#1{\vbox{\hbox to0pt{\vrule width\tableaurule height#1\hss}}}
\def\sqr{\vbox{%
  \phantomhrule\tableaustep
  \hbox{\phantomvrule\tableaustep\kern\tableaustep\phantomvrule\tableaustep}%
  \hbox{\vbox{\phantomhrule\tableauside}\kern-\tableaurule}}}
\def\squares#1{\hbox{\count0=#1\noindent\loop\sqr
  \advance\count0 by-1 \ifnum\count0>0\repeat}}
\def\tableau#1{\vcenter{\offinterlineskip
  \tableaustep=\tableauside\advance\tableaustep by-\tableaurule
  \kern\normallineskip\hbox
    {\kern\normallineskip\vbox
      {\gettableau#1 0 }%
     \kern\normallineskip\kern\tableaurule}%
  \kern\normallineskip\kern\tableaurule}}
\def\gettableau#1{\ifnum#1=0\let\next=\null\else
\squares{#1}\let\next=\gettableau\fi\next}

\newcommand{\figref}[1]{Fig.~\protect\ref{#1}}
\title{\Huge{\boldmath A toy model for the quark pole mass}}

\author[a]{Marcos Mari\~no}
\author[b]{and Maximilian Schwick}

\affiliation[a]{D\'epartement de Physique Th\'eorique et Section de Math\'ematiques\\
Universit\'e de Gen\`eve, Gen\`eve, CH-1211 Switzerland}

\affiliation[b]{Albert Einstein Center for Fundamental Physics, ITP\\ University of Bern, Bern, CH-3012 Switzerland}

\abstract{We study the pole mass for a heavy quark probe in the Gross--Neveu model, at the first non-trivial order in the $1/N$ expansion. We show that this mass suffers from a perturbative renormalon ambiguity structurally similar to the one in QCD. The exact large $N$ solution leads however to an explicit expression for the pole mass which can be decoded as a trans-series, and we find that non-perturbative corrections cure the renormalon ambiguity. We also calculate the self-energy of the heavy quark beyond perturbation theory by using the OPE method with condensates, and show that the results are in agreement with the large $N$ solution. The non-perturbative corrections to the pole mass turn out to be determined to a large extent by the condensate corrections to the self-energy. However, there is a class of corrections that has to be reorganized due to on-shell effects and cannot be obtained from the conventional condensate expansion. We discuss possible implications of this toy model for QCD.}    

\begin{document}
\maketitle
\flushbottom

 \sectiono{Introduction} 
 
 The quark pole mass in QCD is a well-defined quantity in perturbation theory 
 \cite{tarrach,kronfeld}, as it is gauge-independent and infrared-finite. However, as in 
 many other situations, the resulting perturbative series is expected to be factorially divergent 
 due to renormalon effects. This has been verified in the large $N_f$ approximation, by considering a heavy quark with a 
 mass $m$ in a theory with $N_f-1$ massless quarks. The leading renormalons can then be obtained from bubble diagrams \cite{beneke-braun2, shifmanetal}. In heavy quark theory it is expected that the corresponding renormalon ambiguities cancel when calculating physical observables. For example, in the computation of masses of mesons, the ambiguity in the pole mass should be cancelled by a similar ambiguity in the binding energy (see e.g. \cite{beneke, beneke-pole} for a review and references on these issues). 
 
 In observables with an OPE, the perturbative series can be upgraded to a trans-series by using the 
 OPE with condensates. Renormalons in perturbation theory are then expected to be cancelled by ambiguities 
 in the operator condensates \cite{parisi-renormalons, svz, david2}. In contrast to this well-known scenario, it has 
 been argued that the renormalon in the quark pole mass cannot be related to local condensates \cite{shifmanetal}. However, the idea that 
 the quark pole mass gets non-perturbative corrections from condensates in the OPE was already put 
 forward in \cite{politzer} and predates the development of the SVZ sum rules. This idea was further developed in a series of 
 papers \cite{est, elias2, elias3, elias4} (see also \cite{pdr}), which proposed a self-consistent determination of 
 the quark pole mass by using the OPE of the self-energy. In this approach, the first non-perturbative correction 
 to the $\overline{\text{MS}}$ mass $m$ in QCD, at leading order in the coupling, is schematically of the form 
 \be
 {\langle \overline \bpsi \bpsi \rangle_c \over m^2}, 
 \ee
 where $\langle \overline \bpsi \bpsi \rangle_c$ is the quark condensate 
 (see \cite{elias3} for precise formulae). 
 
 Given these different and sometimes conflicting approaches to the problem, it would be interesting to study the 
 non-perturbative structure of the quark pole mass in a simpler model 
 where one might have better analytic control. An obvious candidate is the Gross--Neveu (GN) 
 model \cite{gross-neveu}, which can be regarded as a toy model for the quark 
 sector of QCD. This model is exactly solvable order by order in the $1/N$ expansion, and observables at each order are given by actual 
 functions of the coupling. These functions contain detailed information about both the perturbative 
 series and its condensate corrections. Therefore, they can be used to perform precision 
 tests of the trans-series obtained in the SVZ method, as it was done in \cite{mm-condensates} building on previous work 
 \cite{david2,bbk} (see \cite{liu, sss, liu-on,anatomy,lm-nlsm} for related works).

To study the quark pole mass in this framework, we consider a GN model with $N$ massless Dirac 
quarks and a ``heavy" quark. To simplify our calculations, we work in the probe 
approximation in which quantum effects are only due to massless quarks. At NLO in the 
$1/N$ expansion, the corrections to this approximation do not affect the renormalon problem we are studying here. 
Although some aspects of the underlying physics are very different (for example, elementary quarks 
are not confined), the GN model clarifies some of the issues raised by the quark pole mass in QCD. 
We first show that, in perturbation theory and at the first non-trivial order in the $1/N$ expansion, the 
quark pole mass has renormalon singularities which are structurally similar to those appearing in QCD. However, 
in the exact large $N$ calculation, the quark pole mass is a well-defined function of the renormalized 
$\overline{\text{MS}}$ mass $m$ and the dynamical scale of the theory. When $m$ is large as 
 compared to the dynamical scale, this function asymptotes to the perturbative result, and it can be 
 decoded as a trans-series with an infinite series of non-perturbative corrections which cure 
 the perturbative renormalon ambiguities. 
 
We next ask what is the origin of these non-perturbative corrections, and to do this we 
first calculate the self-energy of the heavy quark probe in the exact $1/N$ expansion. 
We verify that the first two corrections can be reproduced from two-quark and four-quark 
condensates, generalizing in this way the result of \cite{mm-condensates}. We show that 
some of the non-perturbative corrections to 
the quark pole mass can be obtained in a straightforward manner from 
the condensate contributions to the self-energy. However, there is a class of corrections that 
has to be reorganized in a non-trivial way in order to obtain the correct result, as one can read 
off from the exact large $N$ result. Therefore, the physics of these corrections is still related to 
condensates, but the usual trans-series appearing in the SVZ method, organized by operators 
of increasing dimension at fixed $p^2/m^2$, cannot be used in its conventional form to calculate 
them. 

This paper is organized as follows. In section \ref{sec-pert} we study the 
quark pole mass in perturbation theory and we demonstrate the presence of IR renormalons 
similar to the ones appearing in QCD. In section \ref{sec-np-pole} we use the exact large $N$ method 
to find the quark pole mass at the first non-trivial order in $1/N$, and we decode it as a trans-series by 
using the method of \cite{bbk}. We show explicitly how renormalons cancel against 
non-perturbative corrections, and we extend the trans-series presentation to the full 
self-energy. In section \ref{sec-cond} we generalize the condensate calculation of 
\cite{mm-condensates} to a heavy quark probe, and we verify that the very first two corrections 
obtained in the exact large $N$ method are due to two-quark and four-quark condensates. We 
then show that the self-energy trans-series 
has two pieces: one of them can be evaluated on-shell, while another requires a significant 
reorganization in order to match the result for the quark pole mass. We conclude 
in section \ref{sec-conclusions} with comments and prospects for future work. We have 
included two appendices. The first one lists useful results for Feynman integrals, and the 
second one summarizes the techniques of Palanques--Mestre and Pascual to calculate bubble 
diagrams \cite{pmp}, which are extensively used in the paper. 
   
\sectiono{The quark pole mass in perturbation theory}
\label{sec-pert}
%The idea of using large $N$ methods to test the SVZ method was first proposed in the context of the two-dimensional %non-linear sigma model goes back to \cite{david2, svz-report}. 

%We will now explore toy models of renormalons in QCD based on the Gross--Neveu model \cite{gross-neveu}. 

%\subsection{Review of mass renormalization and the pole mass} 

We will consider the GN theory in two dimensions with $N+1$ Dirac fermions. There are $N$ massless fermions $\boldsymbol{\psi}=(\psi_1, \psi_2,\cdots, \psi_{N})$, and a ``heavy quark" $\chi$ of bare mass $m_0$ (in general, bare parameters will be denoted by a $0$ subindex, like $m_0$ and $g_0$ in (\ref{Lsigma})). We will realize the model with an auxiliary scalar field $\sigma$, so that the Lagrangian density is given by 
\be
\label{Lsigma}
\CL_\sigma= \ri  \overline{\boldsymbol{\psi}} \cdot \slashed{\partial} \boldsymbol{\psi}+ \ri  \overline{\chi}\slashed{\partial} \chi-m_0 \overline \chi  \chi-{1\over 2 } \sigma^2 +{\sqrt{g_0}}  \sigma \overline{\boldsymbol{\psi}} \cdot \boldsymbol{\psi}+{\sqrt{g_0}}  \sigma \overline{\chi} \chi. 
\ee
We will work in the quenched probe approximation for the heavy quark, in which loop effects are only due to the $N$ light fermions. We will also work in the large $N$ expansion, in which the 't Hooft coupling constant
\be
\label{thooft-par}
\lambda= {g N \over \pi}
\ee
is kept fixed. In the quenched approximation, the beta function is the one in the theory with $N$ massless fermions. It is useful to introduce the beta function for the 't Hooft coupling
\be
\label{betal}
\beta_\lambda (\lambda; \epsilon)= {N \over \pi} \beta(g; \epsilon)=-\epsilon \lambda+\beta_\lambda (\lambda),
\ee
where $\beta(g; \epsilon)$ is the conventional beta function. The function $\beta_\lambda (\lambda)$ has a $1/N$ expansion at fixed 't Hooft coupling given by
\be
 \beta_\lambda (\lambda)=\sum_{j \ge 0}  \beta^{(j)}_\lambda (\lambda)  N^{-j}, 
\ee
and similarly for  $\beta_\lambda (\lambda ;\epsilon)$. We note that 
\be
\label{leading-beta}
\beta_\lambda^{(0)}(\lambda;\epsilon)= -\epsilon \lambda- \lambda^2, \qquad \beta_\lambda^{(j)}(\lambda;\epsilon)=\beta_\lambda^{(j)}(\lambda), \quad j \ge 1. 
\ee

We will consider the heavy quark pole mass and its self-energy up to (and including) order $1/N$. Perturbative calculations will be done in the ${\overline{\text{MS}}}$ scheme. The bare propagator of the heavy quark will be denoted simply by $S(p)$ (the dependence on $m_0$ and other parameters is implicit). The (bare) self-energy of the heavy quark is defined in terms of the (bare) inverse propagator as 
\be
S^{-1}(p)= \slashed p -m_0 -\Sigma(p^2, m_0^2). 
\ee
 It is conventional to write the self-energy as 
\be
\label{sigma-dec} 
\Sigma(p^2, m_0^2)= \slashed p \Sigma_p (p^2, m_0^2) - m_0 \Sigma_m (p^2, m_0^2) = m_0 \Sigma_1 (p^2, m_0^2)+ \left( \slashed p- m_0 \right) \Sigma_2 (p^2, m_0^2),  
\ee
so that 
\be
\Sigma_1= \Sigma_p- \Sigma_m, \qquad \Sigma_2= \Sigma_p. 
\ee
The renormalized or ${\overline{\text{MS}}}$ mass $m$ is defined as
\be
m_0= Z_m m = Z^{-1} \widetilde Z_m m, 
\ee
where $Z$ is the renormalization constant of the quark field. 
%The renormalized propagator is defined as 
%
%\be
%S^{-1}_R (p)=  \slashed p- m -\Sigma_R = \slashed p \left( 1- \Sigma_{2 R} \right) - m \left(1+  \Sigma_{1 R} - \Sigma_{2 %R} \right),  
%\ee
%
%which we will write as 
%
%\be
%S^{-1}_R(p)=\left( 1- \Sigma_{2 R} \right) \left\{ \slashed p - m \left(1+ {\Sigma_{1R} \over 1- \Sigma_{2 R}} \right) \right\}. 
%\ee
%
%In terms of bare quantities we have 
%
%\be
%S^{-1}_R(p) =Z S^{-1}(p)= Z \left( 1- \Sigma_{2 } \right) \left\{ \slashed p - m_0 \left(1+ {\Sigma_{1} \over 1- \Sigma_{2 }} %\right) \right\}, 
%\ee
%
%where $Z$ is the field renormalization. It follows that 
%
%\be
%1-\Sigma_{2R}= Z\left( 1-\Sigma_2 \right), 
%\ee
%
%or, equivalently, 
%
%\be
%1-\Sigma_{pR}= Z\left( 1-\Sigma_p \right). 
%\ee
%
%Similarly, we find 
%
%\be
%Z_m = Z^{-1} \widetilde Z_m, 
%\ee
%
%where 
%
%\be
%\widetilde Z_m= {1- \Sigma_{mR} \over 1-\Sigma_m}, 
%\ee
%
%in terms of $\Sigma_m=\Sigma_2-\Sigma_1$. Equivalently, 
%
%\be
%1-\Sigma_{mR}=\widetilde Z_m \left( 1-\Sigma_m \right). 
%\ee
%
In the model we are considering, the self-energy has a $1/N$ expansion of the form 
\be
\Sigma= \sum_{\ell \ge 1} \Sigma^{(\ell)} N^{-\ell}, 
\ee
i.e. it starts at order $1/N$. The mass and field renormalization constants also have a $1/N$ expansion: 
\be
Z=1 + {1\over N} Z^{(1)}+ \cdots, \qquad  Z_m=1 + {1\over N} Z_m^{(1)}+ \cdots,
\ee
and it is easy to check that
\be
Z^{(1)}=[\Sigma^{(1)}_2]^{\rm div}, \qquad Z^{(1)}_m=-[\Sigma^{(1)}_1]^{\rm div}.  
\ee
In particular
\be
\label{ren-1N}
\Sigma^{(1)}_{1R}=\Sigma^{(1)}_{1}-Z^{(1)}+\widetilde Z_m^{(1)}=\Sigma^{(1)}_{1}+ Z_m^{(1)}. 
\ee

The pole mass of the heavy quark, $m_p$, is defined as the pole of the 
renormalized propagator, i.e. it is defined by the equation 
\be
\slashed p- m - \Sigma_R (p^2,m^2)\bigl|_{\slashed p=m_p}=0. 
\ee
By exploiting again the $1/N$ expansion, we find, 
\be
m_p= m \left( 1+  {1\over N} \Sigma^{(1)}_{1R} (m^2, m^2)+ \cdots\right).  
\ee

\begin{figure}[!ht]
\leavevmode
\begin{center}
\includegraphics[height=3cm]{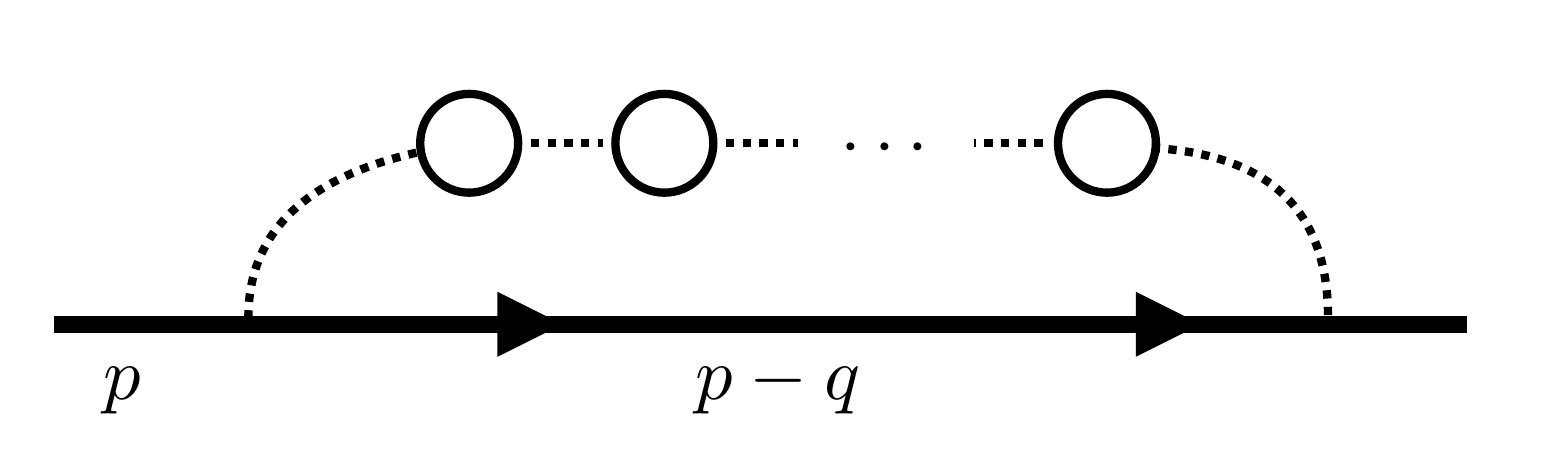}
\end{center}
\caption{Diagrams contributing to the $1/N$ correction to the heavy quark propagator. The thick lines are massive quark propagators, the thin lines are massless quark propagators, and the dashed lines are $\sigma$ propagators.}
\label{mass-bubbles-fig}
\end{figure}

The calculation of the self-energy of the heavy quark at leading order in $1/N$ is very similar to what was done in \cite{mm-condensates}. The relevant diagrams are of the form shown in \figref{mass-bubbles-fig}, in which the massive quark exchanges a sigma particle with insertions of massless fermion loops. They are of course very similar to the relevant diagrams in QCD \cite{beneke-braun2, shifmanetal}. The massless fermion loops in \figref{mass-bubbles-fig} involve the polarization loop 
\be
\label{pint}
\ri \Pi (p^2)=\ri \int {\rd^d q \over (2 \pi)^d} \tr \left[ {1\over \slashed q  ( \slashed p + \slashed q)} \right]= {\mathfrak{g}(\epsilon) \over \pi}  \left(-\frac{p^2}{4\pi}\right)^{-\epsilon/2},
\ee
where
\be
\label{polpre}
\mathfrak{g}(\epsilon)=-\frac{1}{2}\frac{\Gamma(1-\epsilon/2)\Gamma(1+\epsilon/2)\Gamma(-\epsilon/2)}{\Gamma(1-\epsilon)}.
\ee
The self-energy of the heavy quark is then given by the following sum over bubbles, 
\be
\label{sigma-int}
\Sigma^{(1)} (p^2, m^2)= -\ri \sum_{n \ge 1} (\pi \lambda_0)^n \int {\rd^d q \over (2 \pi)^d} {\left(\ri \Pi(q^2) \right)^{n-1} \over \slashed p - \slashed q-m_0}. 
\ee
We note that at this order in $1/N$ we can set $m=m_0$. A standard calculation gives
\be
\label{self-energy}
\ba
&\Sigma_1^{(1)}(p^2, m^2)\\
&=- \pi \ri  \sum_{n \ge 1}  \lambda_0^n \mathfrak{g}^{n-1} \left(-\frac{1}{4\pi}\right)^{-(n-1)\epsilon/2} \left( 2 \CI \left(1, {\epsilon(n-1) \over 2}; p^2, m^2 \right) - \widetilde 
\CI \left(1, {\epsilon(n-1) \over 2}; p^2, m^2 \right) \right), 
\ea
\ee
where the integrals $\CI$, $\widetilde \CI$ are defined in (\ref{mass-res}). They are evaluated on-shell at $p^2=m^2$ by using the expressions in (\ref{on-shell}). To organize the result, we use the approach of \cite{pmp}, which is summarized in Appendix \ref{pmp-trick}. We have 
\be
\Sigma_1^{(1)}(m^2, m^2)=  \sum_{n \ge 1} (Z_\lambda \lambda)^n {F(\epsilon, n \epsilon) \over n \epsilon^n}. 
\ee
The structure function $F(\epsilon,y)$, defined in general in (\ref{eq_structure_function}), reads in 
our case
\be
F(\epsilon,y)=-{1\over 2} \left( {m^2 \over 4 \pi  \nu^2} \right)^{-{y\over 2}}  \left(\frac{\Gamma^2(1-{\epsilon \over 2})\Gamma(1+{\epsilon \over 2})}{\Gamma(1-\epsilon)} \right)^{{y\over \epsilon}-1}\Gamma\left(1+ {y \over 2} \right) \left\{ {2 \Gamma(1-y) \over \Gamma\left(1-{\epsilon+y\over 2} \right)}- {\Gamma(2- y) \over \Gamma\left(2-{\epsilon+y\over 2} \right)} \right\}, 
\ee
where
\be
\nu^2=\mu^2 \re^{\gamma_E - \log(4\pi)}. 
\ee
The divergent part, which can be used to compute $Z_m^{(1)}$, is given by the general formula (\ref{F0div}), where in our case 
\be
\label{F0}
F_0(\epsilon)= F(\epsilon,0)=-{1 \over 2} {1-\epsilon \over 1-\epsilon/2} {\Gamma(1-\epsilon) \over \Gamma^3\left(1-{\epsilon \over 2} \right) 
\Gamma\left( 1+{\epsilon\over2} \right)}. 
\ee
The general expression for the finite part of the sum over bubbles is given in (\ref{finite-part}). It is the sum of a convergent series given by the first two terms, and a factorially divergent series determined by $F(y)= F(0,y)$. 
It is convenient to set $\mu^2=m^2$, so that the perturbative series involves the running coupling constant at the scale set by $m$, which we will denote by $\lambda_{\rm m}$. 
At leading order in $1/N$ it satisfies
\be
\label{lambda-m}
 \re^{-1/\lambda_{\rm m}}\approx {\Lambda\over m}, 
\end{equation}
where $\Lambda$ is the dynamically generated scale in the $\overline{\text{MS}}$ scheme. Then, the structure 
function is 
\be
\label{stru-pole}
F(y)= -{1\over 2}{\Gamma\left(1+ {y \over 2} \right) \Gamma\left(1-y \right) \over \Gamma\left(2-{y\over 2} \right)} = \sum_{n \ge 0} F_n y^n.
\ee
The resulting perturbative series is
\be
m_p-m ={m \over N} \sum_{n \ge 1} r_{n} \lambda_{\rm m}^n +\CO(N^{-2}), 
\ee
where 
\be
r_n= (n-1)! F_{n} +{(-1)^{n-1} \over n}F_{0, n}. 
\ee
$F(y)$ has poles at all positive odd integers and at $y=2$, due to the factor $\Gamma(1-y)$ (poles at even integers greater than $2$ are cancelled by the Gamma function in the denominator). These are the IR renormalons of the 
quark pole mass in this model. The leading factorial growth is controlled by the first IR pole at $y=1$. 

It is not difficult to calculate the full self-energy. The structure functions corresponding to $ \Sigma^{(1)}_{p,m}$ are given, respectively, by
\be
\label{Fp-Sigma}
\ba
F_{p}(\epsilon, y) &=-{1\over 2} \left( {m^2 \over 4 \pi \nu^2} \right)^{-y/2} 
\left( {\Gamma \left(1+\frac{\epsilon}{2}\right) \Gamma^2\left(1-\frac{\epsilon}{2}\right)\over \Gamma (1-\epsilon)}\right)^{\frac{y}{\epsilon}-1} \Gamma \left(1-\frac{y}{2}\right) \Gamma \left(1+\frac{y}{2}\right)\\
   &\times \left(
   _2{\bf{F}}_1\left(\frac{y}{2},\frac{y-\epsilon}{2};1-\frac{\epsilon}{2};\frac{p^2}{m^2}\right)+\left( {y \over 2}-1 \right) \,
   _2{\bf{F}}_1\left(\frac{y}{2},\frac{y-\epsilon}{2};2-\frac{\epsilon}{2};\frac{p^2}{m^2}\right)\right) 
\ea
\ee
and 
\be
\ba
\label{F1-Sigma}
F_{m}(\epsilon, y) &={1\over 2} \left( {m^2 \over 4 \pi \nu^2} \right)^{-y/2} 
\left( {\Gamma \left(1+\frac{\epsilon}{2}\right) \Gamma^2\left(1-\frac{\epsilon}{2}\right)\over \Gamma (1-\epsilon)}\right)^{\frac{y}{\epsilon}-1} \Gamma \left(1-\frac{y}{2}\right) \Gamma \left(1+\frac{y}{2}\right)\\
   &\times     _2{\bf{F}}_1\left(\frac{y}{2},\frac{y-\epsilon}{2};1-\frac{\epsilon}{2};\frac{p^2}{m^2}\right).
   \ea
\ee
As a check, one finds that the wavefunction renormalization constant at order $1/N$, $Z^{(1)}$, is determined by
\be
\label{Fwave}
F_{p,0}(\epsilon)= {\epsilon \over 2} {1\over 2-\epsilon} {\Gamma(1-\epsilon) \over \Gamma^3\left(1-{\epsilon \over 2} \right) 
\Gamma\left( 1+{\epsilon \over2} \right)}, 
\ee
which is the well-known result (see e.g. \cite{mm-condensates}). The structure functions determining the factorial divergence of the self-energy are given by 
\be
\label{H12}
\ba
   F_p(y)&=-{\pi \over 2} \left( {m^2 \over \mu^2} \right)^{-y/2} 
 {(y /2)^2 \over \sin(\pi y/2)}  \, _2F_1\left(1+\frac{y}{2},\frac{y}{2};2;\frac{p^2}{m^2}\right),\\
F_m(y) &={\pi \over 2} \left( {m^2 \over \mu^2} \right)^{-y/2} {y /2 \over \sin(\pi y/2)}  
   {}_2F_1\left(\frac{y}{2},\frac{y}{2};1;\frac{p^2}{m^2}\right).
\ea
\ee
If we go on-shell and set $p=m$, we can evaluate the hypergeometric functions by using Gauss' formula, 
\begin{equation}
\label{gauss}
_2F_1\left(a,b;c;1\right) = \frac{\Gamma(c)\Gamma(c-a-b)}{\Gamma(c-a)\Gamma(c-b)}
\end{equation}
and the combination $F_p(y)-F_m(y)$ reproduces the function $F(y)$ in (\ref{stru-pole}). We note that the evaluation 
(\ref{gauss}) is in principle only valid provided ${\rm Re}(c-a-b)>0$, which in our case means $y<1$. Its expression in terms of Gamma functions provides a meromorphic continuation in $y$, and it relates correctly the perturbative expansions (which are made around $y=0$). However, since in Borel resummations we have to integrate over $y \in (0, \infty)$, the use of Gauss formula beyond perturbation theory should be taken with care. 

The self-energy we have calculated has singularities at even integer values of $y$, as in 
the massless case (where the only singularity occurs at $y=2$ \cite{mm-condensates}). In contrast, the pole mass 
has additional poles at all positive odd integers. This is precisely as in QCD, where going on-shell also introduces 
additional IR singularities. Although the physics of this model is very different from the physics of QCD, the pole mass 
has the same formal properties. 

\begin{figure}[!ht]
\leavevmode
\begin{center}
\includegraphics[height=3cm]{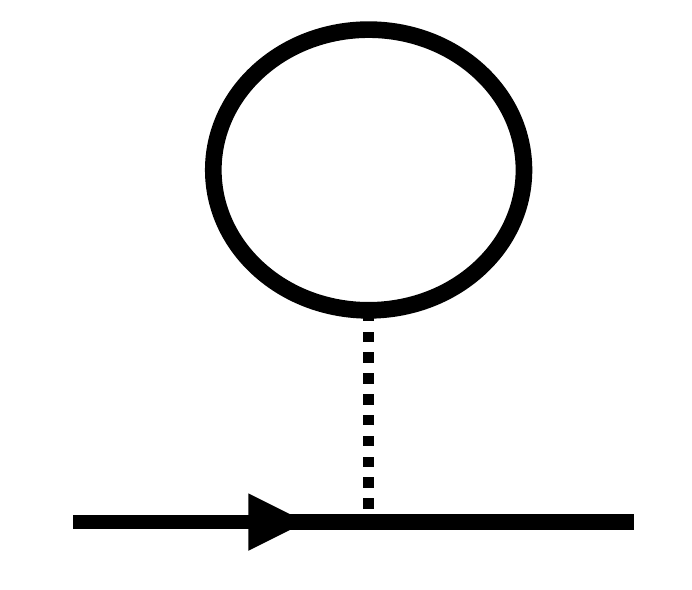}
\end{center}
\caption{At order $1/N$, this diagram in which a heavy quark runs around the loop 
gives the only correction to the quenched approximation for the self-energy.}
\label{heavy-1loop}
\end{figure}

In this section we have considered the perturbative calculation of the self-energy of the heavy quark, at order $1/N$, and in the quenched approximation. At this order in the $1/N$ expansion, the only correction to the quenched approximation is due to a one-loop tadpole diagram with a heavy quark running around the loop, shown in \figref{heavy-1loop}. This contribution is proportional to $m \lambda/N$ and does not affect the renormalon structure we have found. 

\sectiono{The non-perturbative quark pole mass} 
\label{sec-np-pole}
\subsection{The non-perturbative propagator for the heavy quark}
We will now discuss what happens in the exact solution of the theory within 
the $1/N$ expansion. At large $N$ we can integrate out the $N$ massless fermions to obtain 
\be
\CL_{\rm eff}= -{\sigma^2  \over 2 g_0} 
-\ri N \log  {\rm det} \left(  -\ri \slashed \partial - \sigma \right) 
+\overline \chi (\ri \slashed \partial -m_0+ \sigma ) \chi,
\ee
after rescaling $\sigma \rightarrow \sigma/{\sqrt{g_0}}$. The VEV of $\sigma$ is determined by the usual gap equation, so that at large $N$ we have two vacua with\footnote{In \cite{mm-condensates}, the non-perturbatively generated mass $M$ was denoted as $m$.}
\be
\sigma_c=\pm M_0.  
\ee
The heavy quark has now shifted masses $m_0\pm  M_0$ at the two vacua, and we will choose 
the vacuum in which the mass is $m_0+M_0$. We recall that the massive fermion is regarded as a probe, 
and its presence does not affect the large $N$ effective potential for $\sigma$. The only 
ingredient we will need from the large $N$ solution is the propagator of the $\sigma$ field: 
\be
\Delta^{-1}(p;M_0)= {\ri \over 2 \pi} \xi \log\left[ {\xi+1 \over \xi-1} \right], \qquad \xi = {\sqrt{1- {4 M_0^2 \over p^2}}}. 
\ee
The non-perturbative propagator of the heavy quark is given by 
\be
S^{-1}_{\rm NP} (p)= \slashed p - m_0-M_0 -{1\over N} \Sigma^{(1)}_{\rm NP}(p)+ \CO(N^{-2}), 
\ee
where the bare self-energy at order $1/N$ is given by 
\be
 \Sigma^{(1)}_{\rm NP} (p)= \int {\rd^d  k \over (2 \pi)^d} {\Delta(k; M_0) \over \slashed p + \slashed k -m_0-M_0} - 2 \pi \ri  \int  {\rd^d  k \over (2 \pi)^d} {M_0 \over k^2- 4 M_0^2}. 
\ee
This expression can be evaluated as follows. Let us introduce the functions
\be 
\label{AB-mass}
\ba
A(x;c_0)&= {1\over 4 x} \int_0^\infty \left( \xi_y \log\left[ {\xi_y+1 \over \xi_y-1} \right] \right)^{-1} \left(1- {c_0+y-x \over {\sqrt{ (x+y+c_0)^2 - 4 xy }}} \right)\, \rd y, \\
B(x;c_0)&=  {1\over 2} \int_0^\infty \left( \xi_y\log\left[ {\xi_y+1 \over \xi_y-1} \right] \right)^{-1} \left( {1 \over {\sqrt{ (x+y+c_0)^2 - 4 xy }}} +{1 - \xi_y \over 2}\right)\, \rd y,
\ea
\ee
where
\be
 \xi_y= {\sqrt{1 +{4 \over y}}}
\ee
and  
\be
c_0=\left( {m_0 \over M_0}+1 \right)^2.
\ee
These functions are defined by convergent integrals, and when $m_0=0$, i.e. $c_0=1$, one recovers the functions $A,B$ introduced in \cite{cr-dr} and further studied in \cite{mm-condensates}. 
One then finds, 
\be
\ba
\Sigma^{(1)}_{\rm NP} =& - \slashed p A\left(-{p^2 \over M_0^2};c_0 \right) 
- (m_0 + M_0) \left(B\left(-{p^2 \over M_0^2};c_0 \right) -  {1\over 2} \int_0^{\infty} \left( \xi_y \log\left[ {\xi_y+1 \over \xi_y-1} \right] \right)^{-1} {1 - \xi_y \over 2}\, \rd y \right) \\
&-{M_0 \over 2} \int_0^\infty {\rd y \over y+4}. 
\ea 
\ee
This expression is divergent, as expected, and needs to be renormalized. We will use the sharp momentum (SM)
cutoff regularization of \cite{cr-dr,cr-review}: we introduce an upper UV cutoff $\Lambda_{\rm UV}$ in the integrals, we expand the integrand at $k \to \infty$, and we subtract the terms that lead to a divergence. The integral over the subtracted terms 
requires a lower cutoff $\mu^2$ to avoid singularities at $k=0$. As usual, we introduce renormalized masses, 
\be
m_0= m Z_m, \qquad M_0= M Z_M, 
\ee
where 
\be
Z_m= 1+ {1\over N} Z_m^{(1)}+ \CO(N^{-2}), \qquad Z_M= 1+ {1\over N} Z_M^{(1)}+ \CO(N^{-2}). 
\ee
We note that at this order in $1/N$ there is no renormalization of the wavefunction in the SM scheme. We have, 
\be
\ba
S_{\rm NP}^{-1}(p)&= \slashed p - m-M -{m\over N} Z_m^{(1)}-{M\over N} Z_M^{(1)}+{1\over N} \slashed p A\left( -{p^2 \over M^2};c \right) +{M \over 2N} \int_0^{\Lambda_{\rm UV}^2} {\rd k^2 \over k^2+ 4 M^2} \\
&+{1\over N} (m+M) \left( 
B\left(-{p^2 \over M^2};c \right) -  {1\over 2} \int_0^{\Lambda_{\rm UV}^2/M^2} \left( \xi_y \log\left[ {\xi_y+1 \over \xi_y-1} \right] \right)^{-1} {1 - \xi_y \over 2}\, \rd y \right)+ \CO(N^{-2}), 
\ea 
\ee
where 
\be
\label{cM}
c= \left( {m \over M}+1 \right)^2.
\ee
The values of $Z_m^{(1)}$, $Z_M^{(1)}$ are fixed by the SM regularization procedure. As an example, we note that in this scheme the divergent piece of the integral  
\be
-{1\over 2} \int_0^{\Lambda^2/M^2} \left( \xi_y \log\left[ {\xi_y+1 \over \xi_y-1} \right] \right)^{-1} {1 - \xi_y \over 2}\, \rd y 
\ee
is 
\be
{1\over 2} \int_{\mu^2}^{\Lambda^2} {\rd k^2 \over k^2 \log(k^2/M^2)}, 
\ee
where $\mu$ is the arbitrary lower cutoff mentioned before. Therefore, 
\be
\label{zM1}
\ba
Z_M^{(1)}=&{1\over 2} \int_{\mu^2}^{\Lambda^2} {\rd k^2 \over k^2 \log(k^2/M^2)} +
 {1\over 2} \int_{\mu^2}^{\Lambda^2} {\rd k^2 \over k^2}\\
&= {1\over 2} \left\{ \log \log\left({\Lambda^2 \over M^2} \right)- \log \log \left({ \mu^2\over M^2}\right) + \log \left( {\Lambda^2 \over \mu^2} \right)\right\}.  
\ea
\ee
This is the result obtained in the massless case. For $Z_m^{(1)}$ we find instead,
\be
\label{zmnp1}
\ba
Z_m^{(1)}=&{1\over 2} \int_{\mu^2}^{\Lambda^2} {1 \over k^2 \log(k^2/M^2)} \rd k^2= {1\over 2} \left\{ \log \log\left({\Lambda^2 \over M^2} \right)- \log \log \left({ \mu^2\over M^2}\right) \right\}.  
\ea
\ee
We can now remove the regulator, take the limit $\Lambda_{\rm UV} \to \infty$, and calculate the resulting integrals explicitly. We obtain in this way the renormalized non-perturbative propagator of the probe heavy quark, 
\be
\label{np-self-energy}
\ba
S_{\rm NP, R}^{-1}(p)&= \slashed p - m-M -{m_1 \over N} -{M_1 \over N}+{1\over N} \slashed p A\left( -{p^2 \over M^2};c \right) +{m+M\over N} 
B\left(-{p^2 \over M^2};c \right) + \CO(N^{-2}), 
\ea 
\ee
where 
%
%\be
%\ba
%m_1&=m \left[ {1\over 2} \int_0^{\Lambda^2/M^2} \left( \xi_y \log\left[ {\xi_y+1 \over \xi_y-1} \right] \right)^{-1} {1 - \xi_y \over %2}\, \rd y +{1\over 2} \int_{\mu^2}^{\Lambda^2} {\rd k^2 \over k^2 \log(k^2/M^2)} \right], 
%\\
%M_1&=M \left[ {1\over 2} \int_0^{\Lambda^2/M^2} \left( \xi_y \log\left[ {\xi_y+1 \over \xi_y-1} \right] \right)^{-1} {1 - \xi_y \over %2}\, \rd y +{1\over 2} \int_{\mu^2}^{\Lambda^2} {\rd k^2 \over k^2 \log(k^2/M^2)} \right] \\
%&+ M \left[- {1\over 2} \int_{0}^{\Lambda^2/M^2} {\rd y \over y+4} +{1\over 2} \int_{\mu^2/M^2}^{\Lambda^2/M^2} {\rd y \over y}  %\right].
%\ea
%\ee
% 
%These expressions are of course finite but $\mu$-dependent. We obtain, explicitly, 
%
\be
\label{m11}
\ba
{m_1 \over m}&= -{1\over 2} \left( \gamma_E + \log \log(\mu^2/M^2)\right), \\
 {M_1 \over M}&=-{1\over 2} \left( \gamma_E + \log \log(\mu^2/M^2)+ \log(\mu^2/4M^2) \right). 
 \ea
\ee
The result for $M_1/M$ is the one in the massless theory \cite{cr-dr}. 
We can now compute the non-perturbative pole mass from this propagator, as a series in $1/N$. A simple calculation gives 
\be
m_p=m_p^{(0)}+ {m_p^{(1)} \over N}+ \CO(N^{-2}), 
\ee
where 
\be
\label{mp1N}
\ba
m_p^{(0)}&=m + M, \\
m_p^{(1)}&=m_1+ M_1-(m+M) \left( A\left( -c ;c \right) +B\left(-c;c \right) \right). 
\ea
\ee

\subsection{Trans-series for the pole mass}

After renormalization, both the propagator and the pole mass of the heavy quark computed 
in the exact large $N$ method are actual functions of the renormalized mass $m$ and the 
non-perturbative scale $M$. We would like to compare these exact expressions with the perturbative results 
obtained in section \ref{sec-pert}. We expect that, as in previous examples 
\cite{bbk,mm-condensates, anatomy}, the asymptotic expansion of the exact result at weak 
coupling will reproduce the perturbative results, after matching the schemes. As we will see, 
this is precisely the case. We will in fact obtain a stronger result: the 
exact functions obtained in the exact large $N$ method are Borel-resummed trans-series. The first term in 
the trans-series is the perturbative result, and in addition we will find an infinite family of 
exponentially small, non-perturbative corrections which can be determined explicitly. In particular, 
the renormalons in the quark pole mass that we found in section \ref{sec-pert} will be cured by 
these non-perturbative corrections.

Following the pioneering work of \cite{bbk}, the first step in this program is 
to find a Mellin transform representation of the exact results. In this section 
we will consider the simpler case of the pole mass, and in the next section 
we will consider the self-energy. A similar procedure has been worked out in detail in \cite{mm-condensates}, so our discussion will be brief. The building block is the elementary Mellin transform, 
\begin{equation}
\label{mellin-a}
K_a (s,t) = \int_0^\infty \xi^a_y  \left[ \frac{\xi_y-1}{\xi_y+1} \right]^t y^{s-1} \rd y. 
\end{equation}
We will need the explicit values
\begin{align}
K_0(s,t)&= 2t \frac{\Gamma(2s)\Gamma(-s-t)}{\Gamma(s-t+1)}\frac{\sin(\pi(s+t))}{\sin(\pi(s-t))},\\
K_{-1}(s,t)&= \frac{\Gamma(2s+1)\Gamma(-s+t)}{\Gamma(s+t+1)}, 
\end{align}
and in particular 
\begin{equation}
K_0(1,t)=  \frac{2}{t^2-1}, \qquad K_{-1}(1,t)= \frac{2}{t(t^2-1)}.
\end{equation}
By manipulations similar to those in \cite{mm-condensates}, we end up with the expression
\be
A\left(-c;c\right)+B\left(-c;c\right)
=\int_0^\infty 
{K_{-1}(1,t)-K_0(1,t)\over4} \rd t 
%\nonumber\\
%&\quad
+\int_0^\infty\rd t\int_{C}{\rd s\over2\pi\ri}
K_{-1}(-s,t)Q(s;c),
\label{eq:AB-single-contour}
\ee
where the contour $C$ is given by 
\be
C=\sigma+\ri\mathbb R,
\qquad -{1\over2}<\sigma<0,
\ee
and
\be
Q(s;c)=
-{\pi c^s\Gamma(2s+1)\over
2(1+s)\sin(\pi s)\Gamma(s+1)^2}.
\label{eq:Q-single-kernel}
\ee
 In order to make contact with perturbation theory, we first note that 
$m$ can be identified with the renormalized ${\overline{\text{MS}}}$ mass of section \ref{sec-pert}, 
and this is why we have used the same notation. In the perturbative regime, 
\begin{equation}
\label{mhat}
\widehat M={M\over m} 
\end{equation}
is exponentially small. More precisely, 
\be
\widehat M= \re^{-1/\lambda_{\rm m}}+ \CO(N^{-1}), 
\ee
so at this order in $1/N$ we can replace $\widehat M$ by its large $N$ value. We conclude that the perturbative 
regime corresponds to 
\be
\label{c-large}
c \gg 1. 
\ee
In view of the expression (\ref{eq:Q-single-kernel}) this means that we 
should close the contour $C$ to the left, when ${\rm Re}(s)<0$. Once this is done, 
the integral in (\ref{eq:AB-single-contour}) can be computed by residues. 
There are two families of simple poles, and an isolated simple pole. The first family consists of what we will 
call ``moving" or $t$-dependent poles of $K_{-1}(-s,t)$.  
They are located at $s=-t-k$, with $k \in \IZ_{\ge 0}$, and they contribute
\be
{\rm Res}_{s=-t-k}K_{-1}(-s,t)Q(s;c)
=c^{-t-k}{\cal H}(t,k),
\label{eq:H-unified-residue}
\ee
where
\be
{\cal H}(t,k)=
{4^{-t-k}(-1)^k\Gamma(1+2t+2k)\Gamma(t+k)
\Gamma\left({1\over2}-t-k\right)
\over
2\sqrt\pi\Gamma(k+1)\Gamma(1+2t+k)(1-t-k)}.
\label{eq:H-unified}
\ee
The ``fixed" poles occur at $s=-1/2-k$, $k \in \IZ_{\ge 0}$, and they contribute
\be
{\rm Res}_{s=-1/2-k}K_{-1}(-s,t)Q(s;c)
=c^{-1/2-k}{\cal G}(t,k),
\label{eq:G-unified-residue}
\ee
with
\be
{\cal G}(t,k)=
{4^{-1/2-k}(-1)^k\Gamma(k+1/2)\Gamma(2k+2)
\Gamma(t-k-1/2)
\over
2\sqrt\pi\Gamma(k+1)(1/2-k)\Gamma(t+k+3/2)}.
\label{eq:G-unified}
\ee
Finally, there is an extra pole at $s=-1$ which gives 
\be
\label{fixed-res-1}
{\rm Res}_{s=-1} K_{-1}(-s,t)Q(s;c)= -{K_{-1}(1,t) \over 4 c}. 
\ee
We can then organize the integrals in a power series in $1/c$, and we obtain, 
\begin{align}
A\left(-c;c\right)+B\left(-c;c\right)
&=\int_0^\infty\left\{ c^{-t} \CH(t,0)-{1\over 2t (1+t)} \right\}\rd t -{1\over4c} \int_0^\infty K_{-1}(1,t) \,\rd t  \nonumber\\
&+\int_0^\infty
\left( \sum_{k=1}^\infty c^{-t-k}{\cal H}(t,k)+\sum_{k=0}^\infty c^{-1/2-k}{\cal G}(t,k) \right) \rd t.
\label{eq:intermediate-transseries}
\end{align}
As usual with this type of expressions \cite{bbk}, the different terms appearing inside the integral, in the sum in the r.h.s. 
of (\ref{eq:intermediate-transseries}), are separately singular, since they have simple poles. However, these poles 
cancel in the whole sum, due to the identity
\be
{\rm Res}_{t=1/2+\ell}
c^{-t-k}{\cal H}(t,k)
=-{\rm Res}_{t=1/2+\ell}
c^{-1/2-k-\ell}{\cal G}(t,k+\ell),
\qquad k,\ell\geq0, 
\label{eq:pole-cancelation-unified}
\ee
and the additional identities at $t=0,1$  
\be
\ba
&{\rm Res}_{t=0} \left[ c^{-t}{\cal H}(t,0)-{1\over 2t (t+1)} \right]=
{\rm Res}_{t=0}\left[c^{-t-1}{\cal H}(t,1)
-{K_{-1}(1,t)\over 4c} \right]\\
&={\rm Res}_{t=1} \left[ c^{-t}{\cal H}(t,0)-{K_{-1}(1,t)\over 4c}  \right]=0. 
\ea
\label{eq:integer-pole-cancelation}
\ee
After expanding in $1/c$, the individual terms are separately ambiguous and the integrals over $t$ must be done with lateral prescriptions. It is more natural however to use the non-perturbative parameter $\widehat{M}$, so we 
have to expand as well $c^{-u}$ by using e.g. the expression, 
\be
c^{-u/2} = \re^{-u/\lambda_{\rm m}}\left(1+\sum\limits_{n=1}^{\infty}{(-1)^n \over n!}\prod\limits_{j=0}^{n-1}(u+j)\re^{-n/\lambda_{\rm m}}\right). 
\ee
In this way we find an expansion of the form 
\begin{equation}
A\left(-c;c\right)+B\left(-c;c\right)
=\sum_{k\ge 0} \widehat M^k \Xi^\pm_k(\lambda_{\rm m}), 
\label{eq:AB-transseries}
\end{equation}
where 
\be
\label{Xik-def}
\Xi^\pm_k (\lambda_{\rm m})= \int_{\mathcal C_\pm}\rd t\,
\left\{\re^{-2t/\lambda_{\rm m}}b_k(t)+r_k(t)\right\}. 
\ee
In this expression, $\CC_\pm$ are integration contours slightly above or below the positive real axis, respectively, which make it possible to define the integral of each separate summand. Due to the cancellation mentioned above, the full sum is invariant under the choice of contour. The functions $b_k(t), r_k(t)$ can be read from the previous results, and we note that $r_k(t)$ are rational functions of $t$. We have, for example, 
\be
\ba
b_0(t)&= {\cal H}(t,0), \qquad r_0(t)= -{1\over2t(1+t)}, \\
b_1(t)&=-2t{\cal H}(t,0) =
2F(2t),\qquad
r_1(t)={\cal G}(t,0)={1\over2(t^2-1/4)}, 
\ea
\label{brs}
\ee
where $F(y)$ was defined in (\ref{stru-pole}). We also note that, in general, both $b_k(t)$ and $r_k(t)$ are singular at $t=0$, although those singularities cancel in the integrand of (\ref{Xik-def}). 

We now use these results, together with (\ref{mp1N}), to write the $1/N$ correction to the quark pole mass as a laterally Borel resummed trans-series. We have, 
\begin{equation}
m^{(1)}_p=M_1+ m \sum_{k \ge 0} \widehat M^k\mathcal M^\pm_k(\lambda_{\rm m}), 
\label{eq:pole-mass-transseries}
\end{equation}
where 
\be
\ba
\mathcal M^\pm_0&={m_1\over m}-\Xi^\pm_0, \\ 
\mathcal M^\pm_r&=-\Xi^\pm_{r-1}-\Xi^\pm_r,
\qquad r\geq1.
\ea
\label{eq:Mr-from-Phi}
\ee
Let us first look at the leading term $\CM^\pm_0$. The asymptotic expansion of these functions 
should match the perturbative answer found in 
section \ref{sec-pert}, once the corresponding schemes are matched. In order to see this, we write
\be
\mathcal M_0^{\pm}
={m_1 \over m} -\CL (\lambda_{\rm m})+ \int_{\mathcal C_\pm}\rd t\,
\re^{-2t/\lambda_{\rm m}}\mathfrak{m}_0(t),
\label{eq:pole-sector-zero}
\ee
where 
\be
\CL(u)
=\int_0^\infty\left\{
{\re^{-2t/u}\over2t}
-{1\over2t(1+t)}\right\}\rd t =-{1\over2}\left(\gamma_{\rm E} + \log(2) \right)+{1\over 2} \log(u)
\label{eq:I0-pole}
\ee
and 
\be
\mathfrak{m}_0(t)
={F(2t)-F(0)\over t},
\qquad F(0)=-{1\over2}. 
\label{eq:perturbative-B-kernel}
\ee
By looking at (\ref{m11}) we see that the first two terms in the r.h.s. of (\ref{eq:pole-sector-zero}) 
cancel precisely when the lower cutoff in the SM scheme is $\mu=m$, and 
$\mathfrak{m}_0(t) \rd t $ is exactly the Borel kernel $\CB(y) \rd y$ (\ref{borelF}) associated 
to the structure function (\ref{stru-pole}), after setting  $t=y/2$. 

For $k \ge 1$ we find the structure 
\be
\mathcal M^\pm _k
=\int_{\mathcal C_\pm}\rd t\,
\left\{\re^{-2t/\lambda_{\rm m}}\mathfrak{m}_k(t)
+\mathcal R_k(t)\right\}, 
\label{eq:pole-sector-two}
\ee
where $\CR_k(t)$ are rational functions of $t$. It follows from the previous discussion that both 
$\mathfrak{m}_k(t)$ and $\CR_k(t)$ are singular at $t=0$, although singularities cancel. It is then 
convenient to write 
\be
\label{CMk}
\CM^\pm_k (\lambda_{\rm m})= \int_{\mathcal C_\pm}\rd t\,
\left\{\re^{-2t/\lambda_{\rm m}} \left(\mathfrak{m}_k(t)+
{\rho_{k,0} \over t} \right)+\CR_k(t)- {\rho_{k,0} \re^{-2 t/\lambda_{\rm m}} \over t} \right\}, 
\ee
where
\be
\rho_{k,p}= {\rm Res}_{t=p} \CR_k(t). 
\ee
We note that $p$ can be integer or half-integer. 
We find, for the first non-perturbative corrections,
\be
\ba
\mathfrak{m}_1(t)
&=\left({1\over2}-t\right){4^{-t}\over\sqrt\pi}
\Gamma\left({1\over2}-t\right)\Gamma(t-1)
={1-2t \over t} F(2t),\\ 
\mathcal R_1(t)
&={1\over 2t(t+1)}-{1\over2(t^2-1/4)},
\\
\mathfrak{m}_2(t)
&={4^{-t-1}\left(1+t-t^2-4t^3\right)\over\sqrt\pi}
\Gamma\left(-t-{1\over2}\right)\Gamma(t-1)
=F(2t)\left\{ -2-{1\over t}+2t+{3(1+t)\over1+2t}\right\},\\
\mathcal R_2(t)
&={1\over2t(t^2-1)}.
\ea
\label{mrs}
\ee
The integral over the last terms in (\ref{CMk}) can be computed explicitly as (see \cite{mm-condensates})
\be
\label{Hint}
\int_{{\cal C}_\pm} \left( \CR_k(t) - \rho_{k,0} \frac{\re^{-2t/\lambda_{\rm m}}}{t} \right)\rd t =-\rho_{k,0}\log(\lambda_{\rm m})+ \mathfrak{c}_k  \mp \ri\pi \sum_{p \in {\rm P}_k} \rho_{k,p}. 
\ee
In this formula, ${\rm P}_k$ is the set of poles of $\CR_k(t)$, and $\mathfrak{c}_k$ is the principal value integral
\begin{equation}
\mathfrak{c}_k = {\rm{P}} \int_0^\infty \left( \CR_k(t) - \rho_{k,0} {\re^{-2t} \over t} \right)\rd t.
\end{equation}
For $k=1,2$, we find
\begin{align}
\int_{\mathcal C_\pm} \left( \mathcal R_1(t)-{\re^{-2t/\lambda_{\rm m}}\over2t}\right) \rd t
&=-\CL(\lambda_{\rm m}) \pm{\ri\pi\over2},
\label{eq:R1-lateral}\\
\int_{\mathcal C_\pm}\left( \mathcal R_2(t)
+{\re^{-2t/\lambda_{\rm m}}\over2t}\right) \rd t
&=\CL(\lambda_{\rm m})
\mp{\ri\pi\over4}.
\label{eq:R2-lateral}
\end{align}
We conclude that the $1/N$ correction to the pole mass is 
given by the lateral Borel resummation
\be
{m_p^{(1)} \over m} = { M_1 \over m} +  \, s_\pm \left(\Phi_{\rm m}\right)(\lambda_{\rm m} ), 
\ee
where $\Phi_{\rm m}$ is the trans-series  
\be
\Phi_{\rm m}(\lambda_{\rm m} )=\sum_{k \ge 0} \re^{-k/\lambda_{\rm m}} \varphi_{{\rm m},\, k} (\lambda_{\rm m}), 
\ee
and $\varphi_{{\rm m},\, k} (\lambda_{\rm m})$ are factorially divergent series 
\be
\varphi_{{\rm m},\, k} (\lambda_{\rm m})= \sum_{\ell \ge 0} \left( {\lambda_{\rm m} \over 2} \right)^{\ell +1} \widehat{\mathfrak{m}}_k^{(\ell)}(0) -\rho_{k,0}\log(\lambda_{\rm m})+ \mathfrak{c}_k  \mp \ri\pi \sum_{p\in {\rm P}_k} \rho_{k,p}, 
\ee
where 
\be
\widehat{\mathfrak{m}}_k(t)=\mathfrak{m}_k(t)+{\rho_{k,0} \over t}. 
\ee
We find, for the very first orders, 
\begin{align}
\varphi_{{\rm m},\, 0} (\lambda_{\rm m})
&=-{\lambda_{\rm m}\over4}
-{3+\pi^2\over24}\lambda_{\rm m}^2
-{6\zeta(3)+3+\pi^2\over24}\lambda_{\rm m}^3
+\CO(\lambda_{\rm m}^4),
\label{eq:pole-M0-series}\\
\varphi_{{\rm m},\, 1} (\lambda_{\rm m})
&={\gamma_{\rm E}\over2}+{1\over2}\log \left( {2\over\lambda_{\rm m}} \right)
\pm{\ri\pi\over2}
+{\lambda_{\rm m}\over4}
+{3-\pi^2\over24}\lambda_{\rm m}^2
+{3+\pi^2-6\zeta(3)\over24}\lambda_{\rm m}^3
+\CO(\lambda_{\rm m}^4),
\label{eq:pole-M1-series}\\
\varphi_{{\rm m},\, 2} (\lambda_{\rm m})
&=-{\gamma_{\rm E}\over2}-{1\over2}\log \left( {2\over\lambda_{\rm m}} \right)
\mp {\ri\pi\over4}
+{3+\pi^2\over24}\lambda_{\rm m}^2
+{2\zeta(3)-5\over8}\lambda_{\rm m}^3
+\CO(\lambda_{\rm m}^4).
\label{eq:pole-M2-series}
\end{align}
The conditions (\ref{eq:pole-cancelation-unified}) and (\ref{eq:integer-pole-cancelation}) guarantee that 
the renormalon singularities in the perturbative pole mass will be cancelled by terms in the trans-series. 
For example, the first renormalon singularity at $t=1/2$ leads to an ambiguity 
$\mp \pi \ri \re^{-1/\lambda_{\rm m}}/2$ in $\CM_0^\pm$, 
which is cancelled by the opposite term in (\ref{eq:pole-M1-series}). 

\begin{figure}[!ht]
\leavevmode
\begin{center}
\includegraphics[height=6cm]{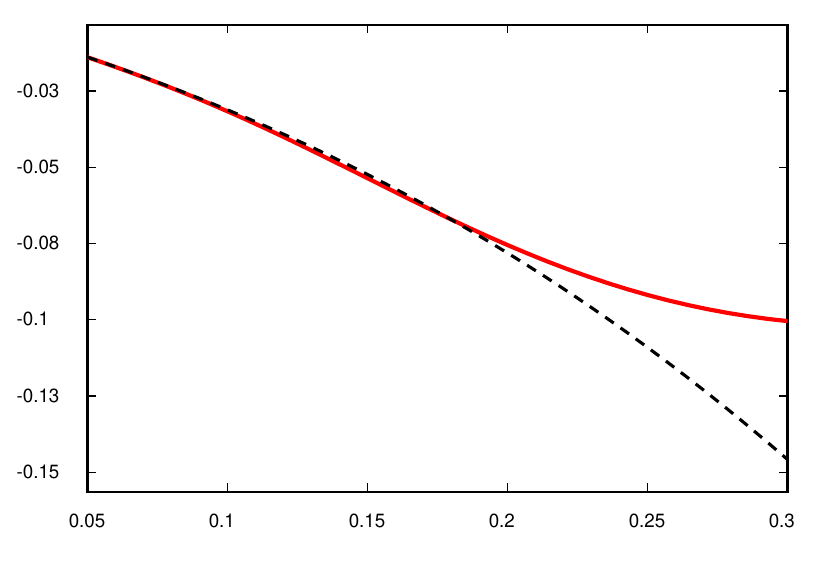}
\end{center}
\caption{The exact function $(m_p^{(1)}-M_1)/m$, as a function of $\lambda_{\rm m}$ (red line), 
and its asymptotic behavior (\ref{eq:pole-M0-series}) as $\lambda_{\rm m} \to 0$, where we kept the first three terms (dashed black line).}
\label{expole-fig}
\end{figure}

We then conclude that, in this model, the pole mass at NLO in $1/N$ is given by an exact function whose asymptotics matches the conventional perturbative expansion, and renormalon ambiguities cancel. In \figref{expole-fig} we show the exact function $(m_p^{(1)}-M_1)/m$ as a function of $\lambda_{\rm m}$, and we compare it to its asymptotic approximation as $\lambda_{\rm m}$ goes to zero. 

\subsection{Trans-series for the self-energy}
\label{subsec:tsse}

It turns out that, with a little bit more effort, we can calculate the trans-series for the full self-energy, generalizing in this way the calculation of \cite{mm-condensates}, which was made for a massless quark. 
We first recall the following Mellin transform \cite{bbk}
\be
f(x, s) = \int_{0}^{\infty}\text{d}y \frac{y^{s+1}}{\sqrt{(x+y+1)^2-4 x y}}={\pi \over \sin (\pi s)} {~}_2 F_1 \left( -s-1, -s-1, 1; -x \right).
\ee
A simple change of variables shows that 
\begin{align}
\int_{0}^{\infty}\text{d}y \frac{y^{s+1}}{\sqrt{(x+y+C)^2-4 x y}} = C^{s+1} f\left(\frac{x}{C}, s\right). 
\end{align}
We define
\be
\mathfrak{f}(s)=c^{s+1} f\left(\frac{x}{c}, s\right)
\ee
as well as 
\be\ba
f_A\left(x, s, c\right) &= \int_{0}^{\infty} \frac{c+y-x}{\sqrt{\left(x+y+c\right)^2-4 x y}}y^{s}\text{d}y= \left(c - x\right) \mathfrak{f}(s-1) +\mathfrak{f}(s),\\
f_B\left(x, s, c\right) &= \int_{0}^{\infty} \frac{y^{s}}{\sqrt{\left(x+y+c\right)^2-4 x y}}\text{d}y =\mathfrak{f}(s-1). \\
\ea
\ee
A simple calculation along the lines of the previous section shows that 
\be
\label{AB-integrals}
\ba
A\left(x, c\right) &=- \frac{1}{4x}\int_{0}^{\infty}\rd t
\int_{C }\frac{\rd s}{2\pi\ri}
K_{-1}(-s, t) f_A(x, s, c), \\
B\left(x, c\right) &= \frac{1}{4}\int_{0}^{\infty}\text{d}t\left(K_{-1}(1,t)-K_{0}(1,t)\right) + \frac{1}{2}\int_{0}^{\infty}\text{d}t\int_{C}\frac{\text{d}s}{2\pi\ri} K_{-1}(-s, t)  f_B(x, s, c).
\ea
\ee
As in the case of the pole mass, these integrals can be computed by residues. First, we have the contribution of ``moving" residues at $s=-t-k$, $k \in \IZ_{\ge 0}$. We will denote
\begin{align}
 - \frac{1}{4x}{\rm Res}_{s=-t-k}K_{-1}(-s, t)  f_A\left(x, s, c\right) &=c^{-k-t} \kappa_{\text{A}}^{(k)}\left(x, c, t\right),\\
\frac{1}{2}{\rm Res}_{s=-t-k}  K_{-1}(-s, t)  f_B\left(x, s, c\right) &=c^{-k-t}  \kappa_{\text{B}}^{(k)}\left(x, c, t\right),
\end{align}
and we find
\be
\label{eq:residuesABintegral}
\ba
\kappa_{\text{A}}^{(k)}\left(x, c, t\right) &={ \pi \over 2} {  (k+t)\Gamma\left(2k+2t+1\right) \over {\Gamma\left(k+1\right)\Gamma\left(k+2t+1\right)\sin (\pi t)}}{}_2 F_1\left(k+t+1, k+t, 2;-\frac{x}{c}\right), \\
\kappa_{\text{B}}^{(k)}\left(x, c, t\right)& ={ \pi \over 2} \, { \Gamma\left(2k+2t+1\right) 
\over \Gamma\left(k+1\right)\Gamma\left(k+2t+1\right) \sin (\pi t) } {}_2 F_1\left(k+t, k+t, 1;-\frac{x}{c}\right). 
\ea
\ee
There is also the contribution of the ``fixed" residues at the poles $s=-k, k\in\mathbb{Z}_{\geq 1}$. We define 
\begin{align}
 - \frac{1}{4x}{\rm Res}_{s=-k} K_{-1}(-s, t)  f_A\left(x, s, c\right) &=\gamma_{\text{A}}^{(k)}\left(x, c, t\right)\\
\frac{1}{2}{\rm Res}_{s=-k} K_{-1}(-s, t)  f_B\left(x, s, c\right) &=\gamma_{\text{B}}^{(k)}\left(x, c, t\right).
\end{align}
By using the relationship
\be
{}_2F_1\left(k,k;1;-u\right)=(1+u)^{-k}P_{k-1}\left( {1- u\over 1+u} \right),
\ee
where $P_k (u)$ are Legendre polynomials, one finds, for $k \ge 2$, 
\begin{align}
\label{gammaAk}
\gamma_{\text{A}}^{(k)}(x,c,t)
&=-{(-1)^k k^2\Gamma(2k)\Gamma(t-k)
\over 2(2k-1)x(c+x)^{k-1}\Gamma(1+k+t)}
\left\{
P_{k-2}\left({c-x\over c+x}\right)
-P_k\left({c-x\over c+x}\right)\right\}. 
\end{align}
The case $k=1$, which is not covered by the formula above, is
\be
\label{gammaA1}
\gamma_{\text{A}}^{(1)}(x,c,t)
={1\over c+x}{1 \over t (t^2-1)}.
\ee
In addition, we find
\be
\label{gammaBk}
\gamma_{\text{B}}^{(k)}(x,c,t)
=-{(-1)^k k\Gamma(2k)\Gamma(t-k)
\over(c+x)^k\Gamma(1+k+t)}
P_{k-1}\left({c-x\over c+x}\right),
\qquad k\geq1.
\ee
Therefore, with the sign convention of (\ref{np-self-energy}), the
renormalized self-energy can be decomposed as
\be
\Sigma^{(1)}_{\rm NP,R}(p)
=\slashed p\,\Sigma^{(1)}_{{\rm NP},p}(p^2,m^2)
-m\,\Sigma^{(1)}_{{\rm NP},m}(p^2,m^2),
\label{eq:self-energy-residue-sum}
\ee
where
\be
\ba
\Sigma^{(1)}_{{\rm NP},p}(p^2,m^2)
&=-\sum_{k \ge 0} c^{-k} \int_{{\cal C}_\pm}  c^{-t} \kappa_{\rm A}^{(k)} (x,c,t)\rd t-\sum_{k \ge 1} \int_{{\cal C}_\pm}   \gamma^{(k)}_{\rm A} (x,c,t) \rd t,\\
\Sigma^{(1)}_{{\rm NP},m}(p^2,m^2)
&=-{m_1+M_1\over m}+ (1+\widehat M)\ \int_{{\cal C}_\pm}  \left( c^{-t} \kappa_{\rm B}^{(0)}(x,c,t)+{K_{-1}(1,t)-K_0(1,t) 
\over 4} \right)  \rd t\\
&+(1+\widehat M)\sum_{k \ge 1} c^{-k} \int_{{\cal C}_\pm}  c^{-t} \kappa_{\rm B}^{(k)}(x,c,t) \rd t 
+ (1+ \widehat M) \sum_{k \ge 1} \int_{{\cal C}_\pm}   \gamma^{(k)}_{\rm B} (x,c,t) \rd t,
\ea
\ee
and $\widehat M$ was defined in (\ref{mhat}). We note that, as before, separate terms 
have singularities on the positive real axis, and this is why we do the integrals along the deformed contours $\CC_\pm$. The singularities cancel in the full sum, as in the case of the pole mass. 

The exact self-energy at order $1/N$ has a trans-series expansion of the form 
  \be
  \label{sigma-ts}
  \Sigma_{ {\rm NP}}^{(1)}(p^2, m^2) \sim M_1+  \sum_{k \ge 0}\widehat M^k  \Sigma_{k}^\pm. 
  \ee
We will sometimes specialize the trans-series $ \Sigma_{k}^\pm$ according to the standard decomposition (\ref{sigma-dec}), and we will write
  \be
  \label{sigmats-dec}
 \Sigma_{k}^\pm=\slashed p \Sigma_{k,p }^\pm-m \Sigma_{k,m}^\pm.
 \ee

Let us first see how to recover the perturbative result from the exact large $N$ calculation, corresponding to the term $k=0$ in (\ref{sigma-ts}). We recall that 
\be
x=-{p^2 \over M^2}=\re^{2/\lambda}+ \CO(N^{-1}), \qquad  {x \over c}= -{p^2 \over \left( m+ M \right)^2}. 
\ee
Here, $\lambda= \lambda(p)$ is evaluated at the scale $p$. In the perturbative limit, $M\to 0$ and 
\be
-{x \over c} \to {p^2 \over m^2}. 
\ee
We will also write
\be
c= {c \over x} \cdot x= -{m^2 \over p^2} \left( 1+{M \over m} \right)^2 \re^{2/\lambda}. 
\ee
Let us define
\be
 H^A_t(z)=t {}_2F_1(t+1,t;2;z), \qquad  H^B_t(z)={}_2F_1(t,t;1;z),\qquad
\label{eq:H-Phi}
\ee
and
\be
\label{ptm}
{\cal P}(t;m)={\pi\over2\sin(\pi t)}
\left(-{m^2\over p^2}\re^{2/\lambda}\right)^{-t}.
\ee
Then, one finds from the explicit formulae above that
\be
\label{kappask0}
\ba
c^{-t} \kappa^{(0)}_{A}\left(-\frac{p^2}{M^2}, c, t\right) &= \CP(t;m) \left( 1+ {M \over m} \right)^{-2t}  H^A_t\left( \frac{p^2}{m^2}\left( 1+ {M \over m} \right)^{-2}\right), \\
c^{-t} \kappa^{(0)}_{B}\left(-\frac{p^2}{M^2}, c, t\right) &= \CP(t;m) \left( 1+ {M \over m} \right)^{-2t} H^B_t\left(\frac{p^2}{m^2}\left( 1+ {M \over m} \right)^{-2} \right). 
\ea
\ee
After setting $t=y/2$ and $p^2=-\mu^2$, we find 
\be
\label{sigmaP}
\ba
\Sigma^\pm_{0,p}& = -\int_{\CC_\pm} \CP(t;m) H^A_t\left(p^2/m^2\right) \rd t,\\
\Sigma^\pm_{0,m} &=\int_{\CC_\pm} \left[ \CP(t;m) H^B_t\left(p^2/m^2\right)-{ \re^{-2 t/\lambda}\over 2 t} \right]
\rd t ,
\ea
\ee
and we obtain the structure functions in (\ref{H12}). 
  
  Let us now work out in some more detail the first non-perturbative corrections to the self-energy, which will be given a condensate interpretation in the next section. We will introduce the notation 
  \be
  z={p^2\over m^2}, \qquad D=z{\rd\over\rd z}, \qquad L=t+D. 
  \ee
  The quantities that appear in the ``moving" residues are functions of 
  \be
c^{-t}
=\left(-{m^2\over p^2}\re^{2/\lambda}\right)^{-t}(1+\widehat M)^{-2t},
\qquad
-{x\over c}={z\over(1+\widehat M)^2}.
\ee
For an arbitrary function $F(z)$ we have the useful expansion
\be
(1+\widehat M)^{-2t}F\left({z\over(1+\widehat M)^2}\right)
=F(z)-2\widehat M LF(z)
+\widehat M^2\left(2L^2+L\right)F(z)+\CO(\widehat M^3).
\label{eq:second-master-expansion}
\ee
 We now study the trans-series, obtained as an expansion in powers of $\widehat M$. There are two contributions to this expansion, up to second order. The first one comes from the expansion of the moving residue with $k=0$, up to order $\widehat M^2$. The second comes from residues with $k \ge 1$. The first set of contributions is
\be
\ba
c^{-t} \kappa_A^{(0)}(t)
&={\cal P}(t;m)\left\{H^A_t-2\widehat M LH^A_t
+\widehat M^2(2L^2+L)H^A_t+\CO(\widehat M^3)\right\},\\
c^{-t} \kappa_B^{(0)}(t)
&={\cal P}(t;m)\left\{H^B_t-2\widehat M LH^B_t
+\widehat M^2(2L^2+L)H^B_t+\CO(\widehat M^3)\right\}.
\ea
\label{eq:second-k0-AB}
\ee
The first non-perturbative correction to the self-energy, at order $\widehat M$, is then given by 
\be
\label{sigmaNP1}
\ba
\Sigma^\pm_{1,p}&= \int_{\CC_\pm} 2 \CP(t;m) t^2{}_2F_1(t+1,t+1;2;z) \rd t ,\\
\Sigma^\pm_{1,m}&=  \int_{\CC_\pm}  \left[\CP(t;m)\left( {}_2F_1(t,t;1;z)
-2 t{}_2F_1(t+1,t;1;z)\right)-{\re^{-2t/\lambda} \over 2t} \right] \rd t \\
&-{1\over 2} \left( \gamma_E + \log(2) \right)+ {1\over 2} \log(\lambda), 
\ea
\ee
 where we used that 
 \be
 LH^A_t(z)=t^2{}_2F_1(t+1,t+1;2;z), \qquad 
L H^B_t(z)=t{}_2F_1(t+1,t;1;z), 
\ee
and in the second line of (\ref{sigmaNP1}) we have used (\ref{eq:I0-pole}). 

Let us now consider the correction at order $\widehat M^2$. In addition to the corresponding terms in (\ref{eq:second-k0-AB}), we should also include the moving residues with $k=1$. 
Since 
\be
c^{-1}
={\widehat M^2\over(1+\widehat M)^2},
\ee
only their leading coefficient in an expansion in $\widehat M$ is required. From (\ref{eq:residuesABintegral}) we find
\be
\ba
c^{-t} \kappa_A^{(1)}(t)
&=2(t+1){\cal P}(t;m)H^A_{t+1}(z)
+\CO(\widehat M),\\
c^{-t} \kappa_B^{(1)}(t)
&=2(t+1){\cal P}(t;m)H^B_{t+1}(z)
+\CO(\widehat M).
\ea
\label{eq:second-k1-moving-AB}
\ee
At this order there is also the contribution of the fixed residues. By using
\be
c+x={1\over\widehat M^2}\left((1+\widehat M)^2-z\right),
\ee
we obtain
\be
\gamma_A^{(1)}(t)=\gamma_B^{(1)}(t)
={\widehat M^2\over(1-z)t(t^2-1)}+\CO(\widehat M^3).
\label{eq:second-fixed-residue}
\ee
The final result for the second order correction to the self-energy is then 
\be
\label{sigmapm2}
\ba
\Sigma^\pm_2&= -\slashed p \int_{\CC_\pm} \left(  {\cal P}(t;m){\cal A}_2(t,z) -{\re^{-2t/\lambda} \over t (1-z) } \right) \rd t -m \int_{\CC_\pm} \left(  {\cal P}(t;m){\cal B}_2(t,z)-{\re^{-2t/\lambda} \over t (1-z)} \right) \rd t\\
&+m^2 {\slashed p +m \over p^2-m^2} \int _{\CC_\pm} \left({1\over t(t^2-1)}+ {\re^{-2t/\lambda} \over t } \right) \rd t,
\ea
\ee
where
\be
\ba
{\cal A}_2(t,z)
&=\left(2L^2+L\right)H^A_t(z)+2(t+1)H^A_{t+1}(z),\\
{\cal B}_2(t,z)
&=\left(2L^2-L\right)H^B_t(z)+2(t+1)H^B_{t+1}(z),
\ea
\label{eq:AB2}
\ee
and the integral in the last line evaluates to 
\be
\int _{\CC_\pm} \left({1\over t(t^2-1)}+ {\re^{-2t/\lambda} \over t } \right) \rd t= -\gamma_E - \log(2) + \log(\lambda) \mp {\pi \ri \over 2}. 
\ee
 Higher order corrections can be computed similarly. 
  
 \sectiono{Condensate corrections and the pole mass}
\label{sec-cond}

\subsection{Condensate corrections to the self-energy}

As we have seen, the exact result for the self-energy that we found in the last section can be decoded into a trans-series, and we have obtained explicit expressions for the first two non-perturbative corrections. 
As in the massless case studied in \cite{mm-condensates}, we expect that these can be recovered in the SVZ approach, i.e. by doing an OPE calculation in the presence of non-zero condensates. We will now verify that this is also the case for the heavy quark probe studied in this paper. The calculation is very similar to the one in \cite{mm-condensates}, to which we refer for further details. As we work in the quenched probe approximation, we will not consider contributions of condensates of the heavy quark.

\begin{figure}[!ht]
\centering
\includegraphics[height=5cm]{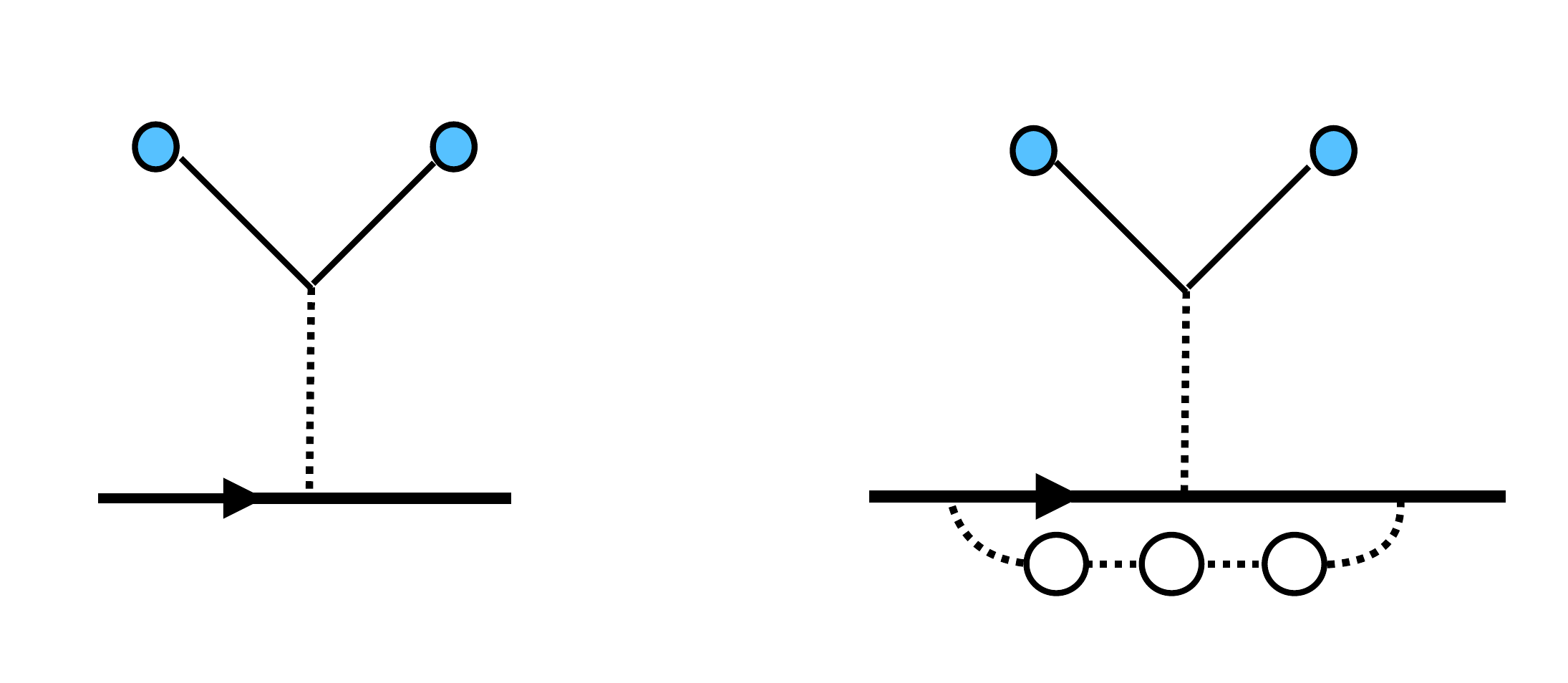}
\caption{The heavy quark propagator with an insertion of a light two-quark condensate.}
\label{2condensate-fig}
\end{figure}

The first condensate correction, as in \cite{mm-condensates} is due to the quark bilinear operator. The corresponding condensate will be denoted by $\langle \overline \bpsi \bpsi \rangle_c$. Composite operators like this have to be renormalized, and we will denote the renormalized composite operator $\CO$ by $[ \CO]$. In the case of the quark bilinear, we have
 \be
 \label{bil-ren}
\overline \bpsi \bpsi = Z_{\overline \bpsi \bpsi}  [\overline \bpsi \bpsi].  
\ee
In addition, renormalization group arguments lead to the equation  \cite{mm-condensates}
\be
\label{nor-ren}
-\frac{\pi\lambda}{N}\,
  \vev{[\overline \bpsi \bpsi]}_c\,
= c(N) \Lambda \left(1
    -\frac1{N}
    \left(\int_0^{\lambda}\frac{\chi(u)-u/2}{u^2}\,\rd u +{1\over 2} \log(\lambda) \right)
    +\CO(N^{-2})
  \right),
  \ee
where
\be
   \label{chiF}
 \chi(u)={u  \Gamma (2+u)\over (2+u) \Gamma^3\left(1+{u \over 2} \right) 
\Gamma\left( 1-{u\over2} \right)}
   \ee
is a renormalization function related to the anomalous dimension of the $\overline \bpsi \bpsi$ operator, and 
\be
\label{cN-expansion}
c(N)= 1+ {c_1 \over N}+ \CO(N^{-2})
\ee
is essentially the numerical value of the condensate in units of $\Lambda$. 

It is easy to see that the only two-quark condensate contribution comes from diagrams of the form shown in \figref{2condensate-fig}. This is the massive counterpart of a class of diagrams already considered in \cite{mm-condensates} (the other two-quark condensate diagrams considered in \cite{mm-condensates} do not contribute in the quenched probe approximation). Their contribution to the self-energy is 
\be
\label{2quark-series}
-\frac{\pi\lambda_0}{N}\,
\vev{\overline \bpsi \bpsi }_c \left\{ 1-{1 \over N} 
  \sum_{n\geq1}
\ri \CI_1(n;p,m)\,\lambda_{0}^{n} \right\} 
\end{equation}
where 
\be
\CI_1 (n;p,m)
  =
  \pi^n\int\frac{\rd^d q}{(2\pi)^d}
  \bigl(\ri\Pi(q^2)\bigr)^{n-1}
  R(p-q), 
\end{equation}
and
\be
R(k)={1\over(\slashed k-m)^2}={1\over k^2-m^2}
+{2m^2\over(k^2-m^2)^2}
+{2m\slashed k\over(k^2-m^2)^2}.
\ee
We introduce now
\be
b={\epsilon(n-1)\over2}.
\ee
We then find, 
\be
\label{eq:massive-condensate-integral}
\ba
\CI_1(n;p,m)=\pi \left( -{1\over 4 \pi} \right)^{-b} \mathfrak{g}^{n-1}\Bigl\{&\CI(1,b;p^2,m^2)+2m^2\CI(2,b;p^2,m^2)\\& +2m\slashed p\left( \CI(2,b;p^2,m^2)
-\widetilde\CI(2,b;p^2,m^2) \right)\Bigr\}, 
\ea
\ee
where $\mathfrak{g}$ is defined in (\ref{polpre}) and the integrals $\CI$, $\widetilde\CI$ are given in Appendix \ref{app-fi}. The sum appearing in (\ref{2quark-series}) is then computed by the following structure functions, 
\be
\label{Fp1-Sigma}
\ba
& {\slashed p \over m} F^{(1)}_{p}(x, y) =-{\slashed p\over 2 m } \left( {m^2 \over 4 \pi \nu^2} \right)^{-y/2} 
\left( {\Gamma \left(1+\frac{x}{2}\right) \Gamma^2\left(1-\frac{x}{2}\right)\over \Gamma (1-x)}\right)^{\frac{y}{x}-1} {\pi y^2/2 \over \sin(\pi y/2)} \\
   &\qquad \times \left(
   _2{\bf{F}}_1\left(1+\frac{y}{2},\frac{y-x}{2};1-\frac{x}{2};\frac{p^2}{m^2}\right)+\left( {y \over 2}-1 \right) \,
   _2{\bf{F}}_1\left(1+\frac{y}{2},\frac{y-x}{2};2-\frac{x}{2};\frac{p^2}{m^2}\right)\right)
\ea
\ee
and 
\be
\ba
\label{Fm1-Sigma}
F^{(1)}_{m}(x, y) &={1\over 2} \left( {m^2 \over 4 \pi \nu^2} \right)^{-y/2} 
\left( {\Gamma \left(1+\frac{x}{2}\right) \Gamma^2\left(1-\frac{x}{2}\right)\over \Gamma (1-x)}\right)^{\frac{y}{x}-1} {\pi y/2 \over \sin(\pi y/2)} \\
   &\times   \left(   _2{\bf{F}}_1\left(\frac{y}{2},\frac{y-x}{2};1-\frac{x}{2};\frac{p^2}{m^2}\right)-y _2{\bf{F}}_1\left(1+\frac{y}{2},\frac{y-x}{2};1-\frac{x}{2};\frac{p^2}{m^2}\right)\right).
   \ea
\ee
We note that $F_p^{(1)}(\epsilon,0)=0$ and 
\be
\label{fm1}
F_m^{(1)}(\epsilon,0)= F_{p,0}(\epsilon)-{\chi(-\epsilon) \over \epsilon}, 
\ee
where $F_{p,0}(\epsilon)$ was introduced in (\ref{Fwave}). 

Let us now consider the 
prefactor appearing in (\ref{2quark-series}). When written in terms of renormalized coupling 
constants and operators, one has to introduce the renormalization factors 
$Z_{\lambda} Z_{\overline \bpsi \bpsi}$. One verifies that these factors, together with the wavefunction renormalization, 
cancel the divergence associated to the structure function (\ref{fm1}), 
exactly like in the massless calculation of \cite{mm-condensates}. Moreover, the 
structure functions $F^{(1)}_{p,m}(y)$ reproduce the Borel integrals appearing in (\ref{sigmaNP1}). After taking into account the change of scheme, just as in \cite{mm-condensates}, we find that the contribution of the condensate diagrams in \figref{2condensate-fig} to the self-energy is of the form 
\be
\label{2condSigma}
 c(N) \Lambda \left\{ 1- {1\over N} \left( {1\over 2} \log(\lambda) + \varphi^{(1)}_m\left( {p^2 \over m^2}; \lambda\right)+ {\slashed p \over m} \varphi^{(1)}_p\left( {p^2 \over m^2}; \lambda\right)\right) \right\}, 
 \ee
 where $\varphi^{(1)}_{m,p}$ are the factorially divergent power series associated to the structure functions $F_{p,m}^{(1)}(y)$ (their general expression is given in (\ref{fborel}).) The term $\log(\lambda)$ originates in the integral in (\ref{nor-ren}), and
 matches the same term appearing in (\ref{sigmaNP1}). The value of $c_1$ 
 in (\ref{cN-expansion}) can be obtained by comparing (\ref{2condSigma}) to (\ref{sigmaNP1}) and using 
 the known relationship between $\Lambda$ and $M$ \cite{fnw1}. One obtains the same value as in \cite{mm-condensates}, as expected. 
 
\begin{figure}[!ht]
\centering
\includegraphics[height=4cm]{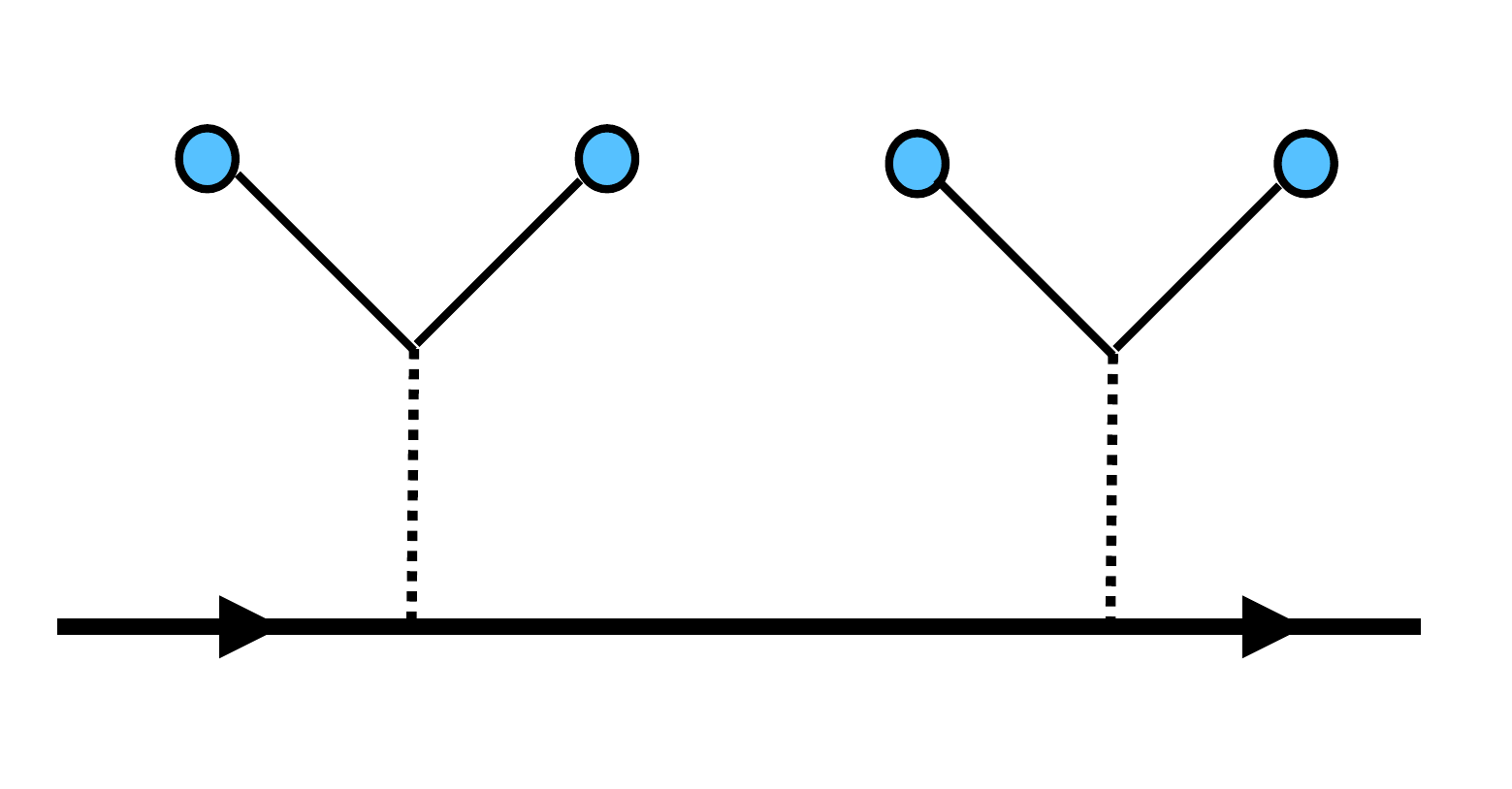}
\caption{A double insertion of a two-quark condensate on the massive fermion line gives a four-quark condensate correction.}
\label{4quark-simple-fig}
\end{figure}

We now consider the corrections due to condensates of higher dimension operators involving the 
$N$ massless fermions. There are two operators of dimension $2$ that one can consider, 
\be
\label{lag-ops}
K = \ri \overline {\bpsi}(x)  \cdot \slashed{\partial} \bpsi (x), \qquad V = g_0 \left(\overline{\bpsi}(x) \bpsi (x) \right)^2. 
\ee
However, their VEVs are related as follows \cite{mm-condensates}, 
\be
\label{KV}
\langle V \rangle_c=-\langle K \rangle_c= \left( 1- {\beta_\lambda(\lambda) \over \epsilon \lambda} \right)  \langle [V] \rangle_c.
\ee
The first equality is a consequence of the equations of motion, and the second gives the renormalization of the operator in terms of the beta function. 

The structure of higher-order condensate corrections was already considered in detail in 
\cite{mm-condensates}, for the massless case. In the quenched approximation we just have to 
generalize the calculation to a heavy quark external line. The simplest contribution is due to the 
diagram shown in \figref{4quark-simple-fig}. As explained in \cite{mm-condensates}, in computing the 
contribution to the self-energy, one has to subtract the square of the first diagram 
in \figref{2condensate-fig}. This is the analogue of considering irreducible diagrams in perturbation theory. 
Let us denote the renormalized two-quark and four-quark condensates by 
\be
\mathcal{T} = - \frac{\pi \lambda}{N} \langle [\overline{\bpsi} \bpsi]\rangle_c, \qquad \CF= {\pi \lambda \over N} \langle [V]\rangle_c. 
\ee
The counterpart of (\ref{nor-ren}) is
\be
\label{4quark-L}
\CF = d(N) \Lambda^2 \exp \left( {1\over N} \left({\beta_\lambda^{(1)} (\lambda) \over \lambda^2}-1 \right) +\CO\bigl(N^{-2}\bigr)\right), 
\ee
where 
\be
\label{dNexp}
d(N)= 1+ {d_1 \over N} + \CO\bigl(N^{-2}\bigr). 
\ee
The fact that $d(N) \approx 1$ at large $N$ is a consequence of large $N$ factorization. Renormalization group arguments imply that \cite{mm-condensates}
 \be
 \CF- \CT^2 ={ \Lambda^2  \over N} \left( d_1-2 c_1 + \log(\lambda) - \int_0^\lambda {\upsilon(u)+u \over u^2} \rd u  \right) + \CO\left(N^{-2}\right), 
 \ee
where
\be
\label{ups-f}
\upsilon(u)=- {u \Gamma(2+ u) \over   \Gamma^3 \left(1+\frac{u }{2}\right) \Gamma \left(1-\frac{u
   }{2}\right)}.
   \ee
Then, the contribution to the self-energy of the ``irreducible" part of the diagram \figref{4quark-simple-fig} is \cite{mm-condensates}
\be
\label{FT}
{ \slashed p+ m\over p^2-m^2} \left\{ \CF -\CT^2 + {\CF \over N}  \int_0^\lambda  {\upsilon (u) \over u(u+ \epsilon)}  \rd u \right\},
\ee
where the second term is the divergent part due to the renormalization of the coupling constant and of the operators. 
As we will see, the terms in (\ref{FT}) involving the renormalization function $\upsilon (u)$ will cancel against divergent and finite parts in other condensate contributions. The remaining part, 
\be
{ \slashed p+ m\over p^2-m^2} { \Lambda^2  \over N} \left( d_1-2 c_1 + \log(\lambda) \right)
\ee
corresponds to the last line in (\ref{sigmapm2}), which came from the first ``fixed" residue. 

\begin{figure}[!ht]
\centering
\includegraphics[height=4cm]{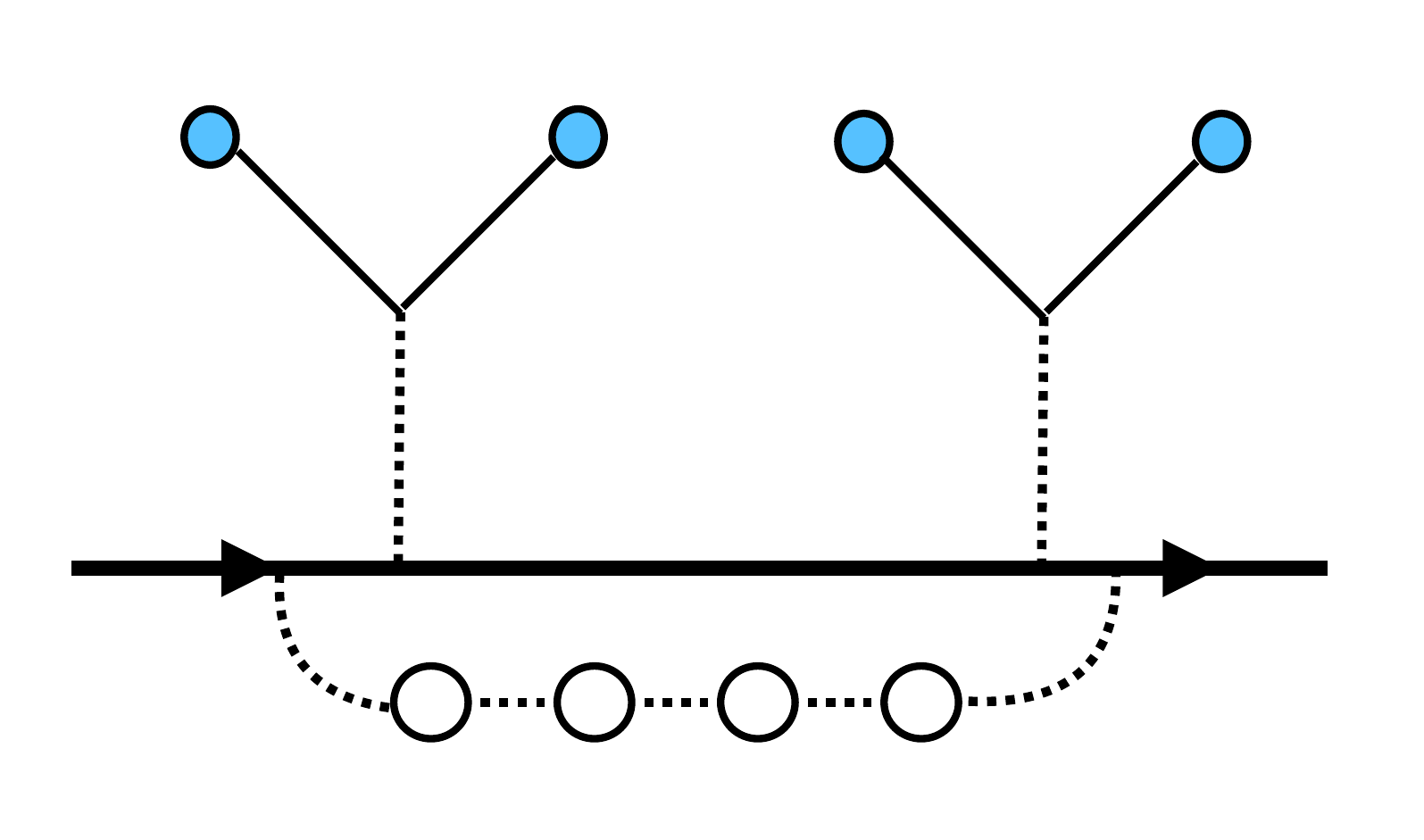}
\caption{A double insertion of a two-quark condensate on the self-energy diagram of \figref{mass-bubbles-fig}.}
\label{4quark-main-fig}
\end{figure}

The second type of contribution is due to the diagrams of the form shown in \figref{4quark-main-fig}, where two two-quark condensates are inserted on the massive fermion line. They lead to the contribution 
\be 
-{\pi \lambda_0 \over N^2} \langle V \rangle_c \sum_{n \ge 1} \ri \CI_2(n) \lambda_0^n, 
\ee
where 
\be
\CI_2(n)=\pi^n\!\int\!\frac{\rd^dq}{(2\pi)^d}
{(\ri\Pi (q))^{n-1} \over (\slashed P-m)^3}
\qquad P=p-q.
\end{equation}
To evaluate this integral we can just exploit the fact that two insertions in the propagating fermion line can be obtained by differentiating the massive propagator twice w.r.t. $m$:
\be
{1\over(\slashed P-m)^3}
={1\over2}{\partial^2\over\partial m^2}{1\over\slashed P-m}.
\label{eq:second-double-insertion}
\ee
Therefore, this contribution is given by  
simply acting twice with the mass derivative on the perturbative result in (\ref{sigma-int}) and (\ref{self-energy}):
\be
\ba
{1\over2}{\partial^2\over\partial m^2}
\left\{{\cal P}(t;m)F(z)\right\}
&={1\over m^2} {\cal P}(t;m)(2L^2+L)F(z), \\
{1\over2}{\partial^2\over\partial m^2}
\left\{ m {\cal P}(t;m)F(z)\right\}
&={1\over m} {\cal P}(t;m)(2L^2-L)F(z).
\ea
\label{eq:second-mass-derivative}
\ee
Taking $F=H^A_t$ and $F=H^B_t$ in the first and second equation, respectively, 
reproduces precisely the first terms of (\ref{eq:AB2}), which come 
from the expansion of the $k=0$ term in (\ref{eq:second-k0-AB}). This result can also be 
obtained by a direct calculation of the diagrams. 

\begin{figure}[!ht]
\centering
\includegraphics[width=14cm]{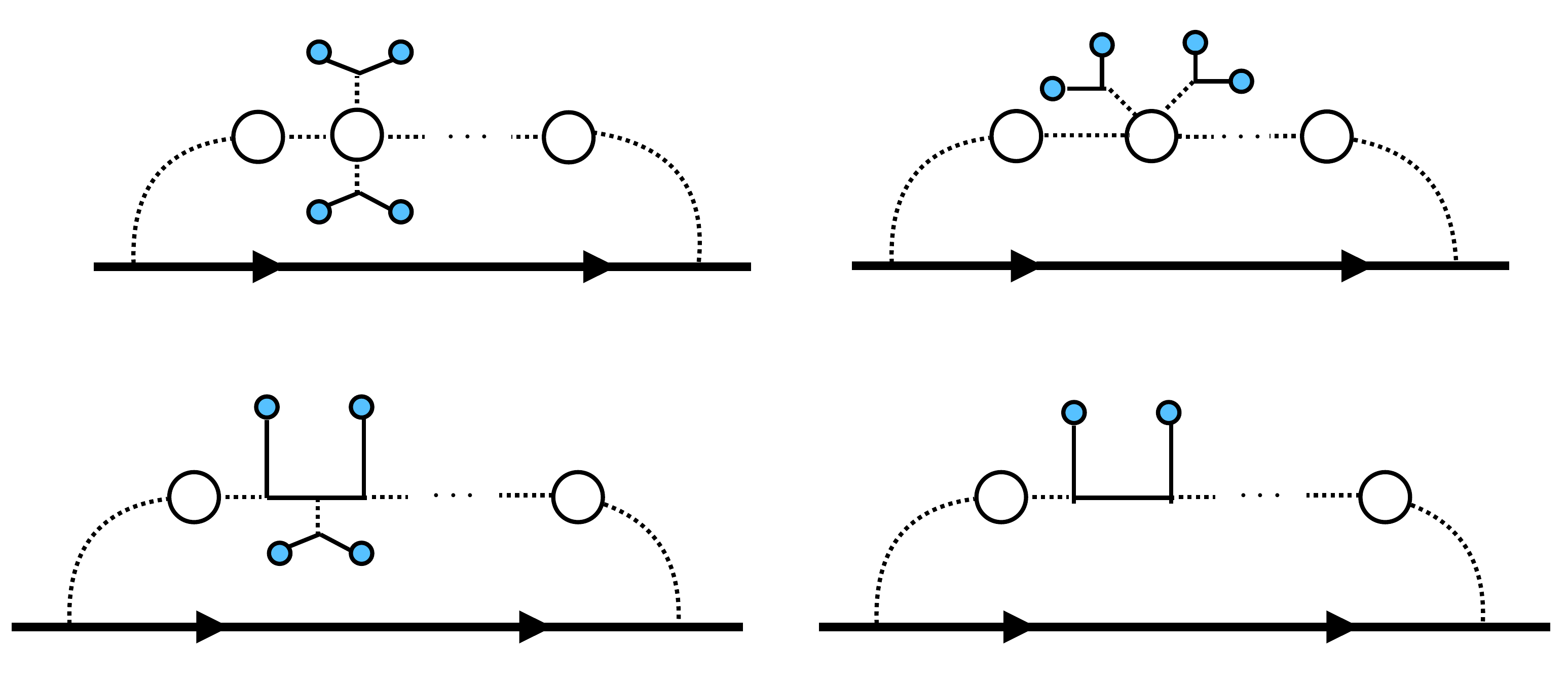}
\caption{The condensate corrections to the chain of bubbles.}
\label{chain-4quark-fig}
\end{figure}

The final contribution to take into account is due to the condensate corrections to the chain of bubbles, or equivalently to the propagator of the $\sigma$ field at order $1/N$. These corrections are due to four types of diagrams. The first two are due to insertions of the condensates in the polarization loops, as shown in the first two diagrams of \figref{chain-4quark-fig}. The third contribution is due to insertion of the condensates in the sigma propagator between the bubbles, as is shown in the third diagram in that figure. Finally, the fourth class of diagrams involves a two-quark condensate insertion in the scalar propagator, but where the quark and antiquark fields are at different points. The derivative expansion of the condensate contributes to the four-quark condensate, as shown in \cite{mm-condensates}.

Let us consider the first two diagrams in \figref{chain-4quark-fig}. As shown in \cite{mm-condensates}, the condensate insertions in the polarization loop lead to a ``decorated" polarization loop given by 
\be
\Pi_{(\bar\psi\psi)^2}(q^2)
=-{2\over q^2}(1-\epsilon)\Pi(q^2). 
\label{eq:second-decorated-polarization}
\ee
Therefore, the contribution of these two diagrams to the self-energy is 
\be
-{\pi \lambda_0 \over N^2} \langle V \rangle_c  \sum_{n \ge 2} \lambda_0^n \ri {\cal D}_1 ^{(n)}, 
\ee
where 
\be
{\cal D}^{(n)}_1
= (n-1)\pi^n \int{\rd^d q\over(2\pi)^d}
{(\ri\Pi(q^2))^{n-2}\ri\Pi_{(\bar\psi\psi)^2}(q^2)
\over \slashed p-\slashed q-m}.
\label{eq:second-D1-start}
\ee
The third diagram gives a contribution of the form 
\be
-{\pi \lambda_0 \over N^2} \langle V \rangle_c  \sum_{n \ge 1} \lambda_0^n \ri {\cal D}_2 ^{(n)}, 
\ee
where 
\be
{\cal D}^{(n)}_2
= 2n \pi^n \int{\rd^d q\over(2\pi)^d}
{(\ri\Pi(q^2))^{n-1}
\over q^2 \left( \slashed p-\slashed q-m \right)}.
\label{eq:second-D2-start}
\ee
Both quantities involve the integral 
\be
\ba
{\cal J}^{(n)}(p,m)&= 
\int{\rd^d q\over(2\pi)^d}
{(q^2)^{-\epsilon(n-1)/2-1}\over \slashed p-\slashed q-m} \\ &=
\slashed p\left\{\CI \left(1,1+{\epsilon (n-1) \over2};p^2,m^2 \right)
-\widetilde\CI \left(1,1+{\epsilon (n-1) \over2};p^2,m^2 \right)\right\}
\\
&+m\CI\left(1,1+{\epsilon (n-1) \over2};p^2,m^2\right),
\ea
\label{eq:second-J}
\ee
and one finds, 
\be
{\cal D}^{(n)}_1+{\cal D}^{(n)}_2
=2\pi \left(1+(n-1)\epsilon\right) \mathfrak{g}^{n-1} \left(-\frac{1}{4\pi}\right)^{-(n-1)\epsilon/2}  {\cal J}^{(n)}(p,m).
\label{eq:D12}
\ee

The final contribution is due to the last class of diagrams shown in \figref{chain-4quark-fig}. If the two massless fermion fields in the two-quark condensate are set at the same point, this diagram is proportional to $\tr\gamma^\mu$ and vanishes. As shown in \cite{mm-condensates}, the first non-zero contribution comes from Taylor expanding the condensate at separated points. In momentum space the separation gives
\be
(y_2-y_3)^\rho
=-{1\over\ri}{\partial\over\partial q_\rho}
\ee
acting on the exponential factor. After integrating by parts one has
\be
{\partial\over\partial q_\rho}\left({q_\rho\over q^2}\right)
={d-2\over q^2}.
\ee
The condensate appearing in this contribution is $\langle K\rangle_c$, but we can use (\ref{KV}) to write it as 
\be
-{\pi \lambda_0 \over N^2} \langle V \rangle_c  \sum_{n \ge 1} \lambda_0^n \ri {\cal D}_3 ^{(n)}, 
\ee
where 
\be
{\cal D}^{(n)}_3=2 \pi {d-2\over d}\,n \mathfrak{g}^{n-1} \left(-\frac{1}{4\pi}\right)^{-(n-1)\epsilon/2} {\cal J}^{(n)}(p,m).
\label{eq:D3}
\ee
After evaluating the integrals appearing in (\ref{eq:second-J}) we obtain the structure functions
\be
\ba
\CR_m (\epsilon,y)
&={1\over  m^2 } \left( {m^2 \over 4 \pi \nu^2} \right)^{-y/2} 
\left( {\Gamma \left(1+\frac{\epsilon}{2}\right) \Gamma^2\left(1-\frac{\epsilon}{2}\right)\over \Gamma (1-\epsilon)}\right)^{\frac{y}{\epsilon}-1} \left(1+ y-\epsilon-{y \over 2- \epsilon} \right) \\
&\times  {\pi y/2 \over \sin(\pi y/2)} 
   {}_2{\bf{F}}_1\left(1+\frac{y}{2}, 1+\frac{y-\epsilon}{2};1-\frac{\epsilon}{2};\frac{p^2}{m^2}\right) 
   \ea
\label{eq:second-structure-B}
\ee
and 
\be
\ba
\CR_{p}(\epsilon,y)
&={1\over  m^2 } \left( {m^2 \over 4 \pi \nu^2} \right)^{-y/2} 
\left( {\Gamma \left(1+\frac{\epsilon}{2}\right) \Gamma^2\left(1-\frac{\epsilon}{2}\right)\over \Gamma (1-\epsilon)}\right)^{\frac{y}{\epsilon}-1} \left(1+ y-\epsilon-{y \over 2- \epsilon} \right) \\
&\times  {\pi y/2 \over \sin(\pi y/2)} \left(1+ {y - \epsilon \over 2} \right) 
   {}_2{\bf{F}}_1\left(1+\frac{y}{2}, 2+\frac{y-\epsilon}{2};2-\frac{\epsilon}{2};\frac{p^2}{m^2}\right).
    \ea
\label{eq:second-structure-A}
\ee
If we set $y=0$ we find the function
\be
\CR_{p}(\epsilon,0)
={\CR}_{m}(\epsilon,0)=-{1\over p^2-m^2} {\upsilon(-\epsilon)\over\epsilon}. 
\label{eq:structure-epsilon}
\ee
This cancels against the contribution of renormalization functions in (\ref{FT}), as in the massless case studied in \cite{mm-condensates}. 
Setting $\epsilon=0$ gives the structure functions
\be
\ba
\CR_p (y)
&={1\over m^2} \left(1+{y\over2}\right){\pi y/2 \over \sin(\pi y/2)}
\left({m^2\over \mu^2}\right)^{-y/2}
H^A_{1+y/2}(z), \\
\CR_m (y)
&={1\over m^2} \left(1+{y\over2}\right){\pi y/2 \over \sin(\pi y/2)}
\left({m^2\over \mu^2}\right)^{-y/2}
H^B_{1+y/2}(z).\\
\ea
\label{eq:second-structure-AB}
\ee
To compare with perturbation theory one now extracts the Borel kernel from these structure functions exactly as in (\ref{borelF}): subtract the value at $y=0$, divide by $y$, set $y=2t$, and include the Jacobian $\rd y=2\rd t$. After setting the subtraction point to $\mu^2=-p^2$, we obtain the integrands
\be
2(t+1){\cal P}(t;m)H^{A,B}_{t+1}(z)
-{\re^{-2t/\lambda}\over(1-z)t}. 
\ee
They reproduce the last terms in the r.h.s. of equations (\ref{eq:AB2}) and the subtraction term in (\ref{sigmapm2}). 
%We note that $\sum_k \gamma^{(k)}_A$, after setting $m=0$ and expanding for $x \gg 1$, reproduces the contributions of the functions $\CH_n(y)$ in \cite{mm-condensates}. 

\subsection{The pole mass from condensates}

So far, the analysis of this paper has established the following facts. First, in this model, the pole mass can be defined 
non-perturbatively from the exact self-energy, in the $1/N$ expansion. It has a trans-series expansion in which the 
usual perturbative renormalon gets cancelled by the ambiguities in the first non-perturbative correction. Second, the 
exact self-energy has a trans-series expansion as well, which is reproduced from condensate calculus up to order $\widehat M^2$. 

These two statements, put together, seem to indicate that the pole mass must be closely related to a condensate calculation. In this final section we study in detail in which sense this is true. As we will see, the trans-series contributions to the pole mass are of two types. The first one can be obtained from the self-energy trans-series, calculated order by order in the condensate expansion. There is however another type of contributions that requires a full reorganization of the self-energy trans-series. 

The first contribution to the pole mass, in the $1/N$ expansion, is simply $m_p^{(0)}= m+M$, where $m$ 
is the $\overline{\text{MS}}$ mass and $M$ the non-perturbative correction, which in the condensate language is due to the first diagram in \figref{2condensate-fig}. Let us now consider the NLO in the $1/N$ expansion. We first introduce the notation
\begin{equation}
 \mathfrak{S}_k^\pm(z)
 =\sqrt z\,\Sigma_{k,p}^\pm(z)-\Sigma_{k,m}^\pm(z).
\end{equation}
Then, at NLO in $1/N$, the pole mass is formally obtained by evaluating the self-energy at $\slashed p= m_p^{(0)}$:
\begin{equation}
\label{mp-cond}
{m_p^{(1)}-M_1 \over m} \sim \sum_{k\geq0}\widehat M^k \mathfrak{S}_k^\pm(z)
   \bigr|_{z=(1+\widehat M)^2}.
\end{equation}
Therefore, the trans-series for the pole mass (\ref{eq:pole-mass-transseries}) can be formally obtained from (\ref{mp-cond}), by expanding the r.h.s. in $\widehat M$. One finds, for the very first orders:
\be
 \label{MsSigmas}
\ba
 \CM_0^\pm&=\mathfrak{S}_0^\pm(1),\\
 \CM_1^\pm&=\mathfrak{S}_1^\pm(1)+2\partial_z\mathfrak{S}_0^\pm(1),\\
 \CM_2^\pm&=
\mathfrak{S}_2^\pm(1)+2\partial_z\mathfrak{S}_1^\pm(1)
 +\partial_z\mathfrak{S}_0^\pm(1)+2\partial_z^2\mathfrak{S}_0^\pm(1). 
 \ea
\ee
Let us examine these equations more carefully. The first equation was tested in section \ref{sec-pert} by using Gauss' formula (\ref{gauss}). To test the second equation, we first note that the momentum-dependent factor in the integrands appearing in the self-energy can be expressed in terms of the coupling at the mass scale:
\be
\left(-{m^2\over p^2}\re^{2/\lambda(p)}\right)^{-t}
=\re^{-2t/\lambda_{\rm m}}.
\label{eq:running-factor-pole-check}
\ee
We now use again the Gauss evaluation of the hypergeometric functions, and obtain
\be
\label{sigmam1}
\ba
\left.\mathfrak{S}_1^\pm(1)\right|_{\rm Borel}&=(t-1)b_0(t),\\
 \left.2\partial_z\mathfrak{S}_0^\pm(1)\right|_{\rm Borel}&=t b_0(t), 
 \ea
\ee
where ``Borel" means that we consider the integrands appearing in the Borel representations of (\ref{sigmaP}) and (\ref{sigmaNP1}), 
and $b_0(t)$ was defined in (\ref{brs}). The sum of the two quantities in (\ref{sigmam1}) gives
\begin{equation}
 (t-1)b_0(t)+t b_0(t)
 =(2t-1)b_0(t)=\mathfrak{m}_1(t).
 \label{eq:first-sum}
\end{equation}
Both contributions in (\ref{sigmam1}) have a very clear interpretation 
in terms of condensates. The first diagram in 
\figref{2condensate-fig} leads to the non-perturbative shift $m \rightarrow m+M$, and hence 
to the second line in \eqref{sigmam1}. The second
diagram in that figure leads to the first line in \eqref{sigmam1}. Finally, we note that the first 
term in $\CR_1(t)$, in (\ref{mrs}), which contributes to $\CM_1^\pm$, follows also from the term 
$(2t (t+1))^{-1}$ in $\mathfrak{S}_1^\pm(1)$. 

Going further, we can test the third equation as follows:
\be
\ba
 \left.\mathfrak{S}_2^\pm(1)\right|_{\rm Borel}
 &=\frac{1-t^3}{1+2t}\,b_0(t), \\
 \left.2\partial_z\mathfrak{S}_1^\pm(1)\right|_{\rm Borel}
 &=-\frac{2t(t^2-1)}{1+2t}\,b_0(t),\\
 \left.\bigl(\partial_z\mathfrak{S}_0^\pm(1)
 +2\partial_z^2\mathfrak{S}_0^\pm(1)\bigr)\right|_{\rm Borel}
 &=-\frac{t(1+t+t^2)}{1+2t}\,b_0(t).
 \ea
 \label{eq:s2}
\ee
Here, the Borel contribution to $\mathfrak{S}_2^\pm(1)$ refers to the integrands in the first line of (\ref{sigmapm2}). 
Their sum gives indeed 
\begin{equation}
 \frac{1+t-t^2-4t^3}{1+2t}\,b_0(t)=\mathfrak{m}_2(t),
 \label{eq:second-sum}
\end{equation}
which is the correct result obtained in (\ref{mrs}). 

So far, in the evaluation on-shell, we have only used the terms in the self-energy which, from the point of view 
of the residue evaluation in section \ref{subsec:tsse} are due to ``moving" residues. 
The contribution of fixed residues turns out to be very different. The last term in 
(\ref{sigmapm2}) gives, at $\slashed p =m+M$, and after multiplication by $\widehat M^2$, a 
contribution of order $\widehat M$. However, it is just the first term in an infinite series. Indeed, if we expand the contribution of fixed residues in powers of $\widehat M$, which is the expansion produced by a fixed-order condensate calculation, we get
\be
\gamma_A^{(k)}(t)+\gamma_B^{(k)}(t)
\sim {\widehat M^{2k}\over(1-z)^{2k-1}}.
\label{eq:fixed-nonuniform-scaling}
\ee
At the pole, $1-z_*=-2\widehat M-\widehat M^2$. We conclude in particular that every fixed residue gives 
a contribution of order $\widehat M$, and an explicit calculation shows that the resulting series is 
divergent. It can be seen that a direct resummation of this series fails to produce the on-shell result. Therefore, one 
needs a more drastic reorganization of this contribution to the self-energy, and the 
way to do it should be clear from the exact answer and its 
Mellin--Barnes representation. Let us consider the sum of the contribution of fixed residues, 
\be
\sum_{k\geq1}\left(\gamma_A^{(k)}(x,c,t)+\gamma_B^{(k)}(x,c,t)\right) =\sum_{k \ge 1} {\rm Res}_{s=-k} 
K_{-1}(-s,t)
\left[-{1\over4x}f_A(x,s,c)+{1\over2}f_B(x,s,c)\right].
\label{eq:fixed-sum-generating}
\ee
Imposing the on-shell condition $x=-c$ inside the integrand, one finds 
\be
-{1\over4x}f_A(x,s,c)+{1\over2}f_B(x,s,c)
=Q(s,c), 
\label{eq:fixed-onshell-generator}
\ee
where $Q(s,c)$ was defined in (\ref{eq:Q-single-kernel}). The pole structure of $Q(s,c)$ is very different from 
the one in the off-shell integrand: as we have seen, it leads to poles at 
half-integer values $s=-1/2 -k$, $k \ge 0$, and at $s=-1$. The Mellin--Barnes integral 
now picks up the residues of these poles. We obtain in this way the sum of the contributions from fixed residues 
 to the pole mass written down in (\ref{eq:G-unified-residue}) and (\ref{fixed-res-1}). In particular, the term $-r_1(t)$ appearing in the first non-perturbative contribution to the integrand of the pole mass originates 
in the reorganization of this tower of residues in the Mellin--Barnes integral. This term is responsible for cancelling the first renormalon ambiguity in the quark 
pole mass, and it does not seem possible to obtain it from an individual local condensate.

The analysis above clarifies the relationship between the pole mass and condensate calculus. The two-quark and 
four-quark condensate calculations reproduce the corrections to the self-energy in detail, as in the massless 
case of \cite{mm-condensates}. As we showed in this section, they can be used to calculate the 
contributions to the pole mass coming from the ``moving" residues, by using Gauss' formula. However, the contribution from 
fixed residues cannot be reproduced by any finite truncation of the ordinary condensate expansion, 
and a reorganization thereof is needed. In fact, we have to first go on-shell in the integrand of the Mellin--Barnes representation, and then pick the new residues. We can summarize the situation by saying that going on-shell and performing the Mellin--Barnes expansion in residues are operations that do not commute. 

\sectiono{Conclusions and outlook}
\label{sec-conclusions}

In this paper we have studied a toy model for the quark pole mass by considering a heavy quark probe in the Gross--Neveu model 
with $N$ massless quarks. The perturbative expansion of the heavy quark pole mass has IR renormalons very similar to 
those found in QCD, and we have found their detailed structure at the first non-trivial order in the $1/N$ expansion. By 
considering the non-trivial large $N$ saddle point of the path integral, one finds an exact result for the pole mass, at NLO in $1/N$, 
which can be decoded as a Borel-resummed trans-series, similar to what is found in other models (see e.g. \cite{anatomy} for a 
similar result for a non-perturbative pole mass in a scalar field theory). 
The trans-series expression shows explicitly that these renormalons are cured by non-perturbative corrections. 

To understand 
the origin of these corrections, we have studied the heavy quark self-energy in the exact large $N$ approach.  
By extending the study in \cite{mm-condensates}, we have shown that its first two non-perturbative corrections to the trans-series can 
be reproduced by the OPE with condensates, as expected. We can then ask to which extent the 
non-perturbative corrections to the quark pole mass originate in the condensates. It turns out that these 
corrections come from two types
of contributions. The first one is obtained directly from the self-energy, by solving the pole mass equation 
and going on-shell. The second one requires, however, a non-trivial reorganization of the usual 
condensate expansion. The existence of the second type of contributions confirms the 
suspicions in \cite{shifmanetal}
that a standard calculation in terms of local condensates cannot reproduce the pole mass correctly. The 
Gross--Neveu model is in this respect particularly interesting, since one of the arguments put forward in \cite{shifmanetal} is that in QCD there are no local operators with the right dimension to produce a non-perturbative correction to the 
mass of order $\Lambda$. In contrast to this, in the Gross--Neveu model the two-quark condensate $\langle\overline \bpsi \bpsi \rangle_c$ has the right dimension, and indeed it leads to the leading shift $m\rightarrow m+ M$.  
In spite of this, the existence of the second type of contributions precludes an obvious derivation of the pole 
mass from the standard condensate expansion.  

There is one obvious lesson of our toy model for QCD: attempts to extract non-perturbative corrections to the 
quark pole mass from the OPE of the self-energy, as in \cite{politzer,est,elias2,elias3,elias4}, might require much more care than exercised in the past, due to on-shell effects of the type found in this paper. Another 
source of complications for that program was already noted in \cite{qcd-propagators}: the non-perturbative 
quark self-energy has contributions due to condensates of non-gauge-invariant operators, which should also be incorporated in the calculation. As a consequence, and in contrast to what is 
found in perturbation theory, it is not clear that the non-perturbative pole mass obtained by any finite truncation of 
the condensate expansion will be gauge-invariant. It is conceivable that a reorganization of the condensate expansion, which in the light of our results could be necessary to address on-shell issues, may also produce a gauge-invariant 
non-perturbative pole mass after including non-gauge-invariant condensates. In our model, the reorganization of the condensate expansion in the on-shell limit was achieved by using the Mellin--Barnes representation of the self-energy. In QCD no analogous, exact non-perturbative representation is available, and it is not clear {\it a priori} how to solve this problem. Perhaps a clever self-consistent approach, as originally envisaged in \cite{est,elias2, elias3, elias4}, might lead to progress on this issue. The toy model presented here would be a useful arena to test such an approach. 

\section*{Acknowledgements}
We would like to thank Yizhuang Liu for useful discussions and comments on the manuscript. 
The work of M. M. is supported in part by the Swiss National Science Foundation, 
subsidy 10007181. The work of M. S. is supported by the Swiss National Science Foundation 
under grant number 200021$\_$219267.

\appendix
\sectiono{Feynman integrals} \label{app-fi}
In this Appendix we collect some explicit results for one-loop Feynman integrals which are used in the paper, see e.g. 
\cite{pascual-tarrach} for some of them. 

We define
\be
\ba
\CI(r, s; p^2)&= \int {\rd^d k \over (2 \pi)^d} {1\over [k^2]^r \left[ (k-p)^2 \right]^s}, \\
\CI^\mu (r, s; p)&= \int {\rd^d k \over (2 \pi)^d} {k^\mu \over  [k^2]^r \left[ (k-p)^2 \right]^s}= p^\mu \widetilde \CI(r, s; p^2). 
\ea
\ee
Our epsilon convention in two dimensions is 
\be
d=2- \epsilon.
\ee
One obtains
\be
\ba
\label{massless-res}
\CI(r, s; p^2)&=-\frac{\ri}{4\pi} \left(-\frac{p^2}{4\pi}\right)^{-\epsilon/2} \frac{1}{(p^2)^{r+s-1}} G(r,s),\\
\widetilde \CI(r, s; p^2)&=-\frac{\ri}{4\pi} \left(-\frac{p^2}{4\pi}\right)^{-\epsilon/2} \frac{1}{(p^2)^{r+s-1}} \widetilde G(r,s),
\ea
\ee
where
 \be
 \ba
 G(r,s)&= \frac{\Gamma( 1 - r - \epsilon/2)\Gamma(1-s-\epsilon/2)\Gamma(r+s-1+\epsilon/2)}{\Gamma(r)\Gamma(s)\Gamma(2-r-s-\epsilon)}, \\
\widetilde G(r,s)&= \frac{\Gamma( 2 - r - \epsilon/2)\Gamma(1-s-\epsilon/2)\Gamma(r+s-1+\epsilon/2)}{\Gamma(r)\Gamma(s)\Gamma(3-r-s-\epsilon)}.
\ea
\ee

In the massive case, we define
\be
\ba
\CI(\alpha, \beta; p^2, m^2)&= \int {\rd^d k \over (2 \pi)^d} {1\over \left[ (k-p)^2-m^2 \right]^\alpha k^{2 \beta}}, \\
\CI^\mu (\alpha, \beta; p^2, m^2)&= \int {\rd^d k \over (2 \pi)^d} {k^\mu \over \left[ (k-p)^2-m^2 \right]^\alpha k^{2 \beta}}= p^\mu \widetilde \CI(\alpha, \beta; p^2, m^2), 
\ea
\ee
which are given by 
\be
\ba
\CI(\alpha, \beta; p^2, m^2)&=\ri  \left( -{p^2 \over 4 \pi} \right)^{d/2} {\Gamma(\alpha+ \beta-d/2) 
\over\Gamma(\alpha) \Gamma(\beta)} {1\over (p^2)^{\alpha+ \beta}} \\
&\times \int_0^1 x^{\alpha-1} (1-x)^{\beta-1} \left[ x(1-x) -{m^2 \over p^2}x \right]^{d/2 -\alpha -\beta}, 
\ea
\ee
as well as 
\be
\ba
\widetilde \CI(\alpha, \beta; p^2, m^2)&=\ri  \left( -{p^2 \over 4 \pi} \right)^{d/2} {\Gamma(\alpha+ \beta-d/2) 
\over\Gamma(\alpha) \Gamma(\beta)} {1\over (p^2)^{\alpha+ \beta}} \\
&\times \int_0^1 x^{\alpha} (1-x)^{\beta-1} \left[ x(1-x) -{m^2 \over p^2}x \right]^{d/2 -\alpha -\beta}, 
\ea
\ee
The integral over $x$ can be evaluated in terms of a hypergeometric function, for arbitrary values of $z=p^2/m^2$. We find, 
\be
\label{mass-res}
\ba
\CI(\alpha, \beta; p^2, m^2)&=\ri  \left( {m^2 \over 4 \pi} \right)^{d/2} (-m^2)^{-\alpha -\beta} 
{\Gamma(\alpha+ \beta-d/2)  \Gamma(d/2-\beta)\over\Gamma(\alpha)} \\
& \times  {}_2{\bf F}_1 \left( \alpha+ \beta -d/2, \beta, d/2; {p^2 \over m^2} \right), \\
\widetilde \CI(\alpha, \beta; p^2, m^2)&=\ri  \left( {m^2 \over 4 \pi} \right)^{d/2} (-m^2)^{-\alpha -\beta} {\Gamma(\alpha+ \beta-d/2)  \Gamma(d/2+1-\beta)
\over\Gamma(\alpha)} \\
& \times ~_2 {\bf F}_1 \left( \alpha+ \beta -d/2, \beta, d/2+1; {p^2 \over m^2} \right),
\ea
\ee
where 
\be
{}_2 {\bf F}_1 \left( a,b, c; z \right)={1\over \Gamma(c)} {}_2 F_1 \left( a,b, c; z \right)
\ee
is the regularized hypergeometric function. 

We are particularly interested in the on-shell value $p^2=m^2$, or $z=1$. In this case, one has
\be
\label{on-shell}
\ba
\CI(\alpha, \beta; m^2, m^2)&=\ri \left( {m^2 \over 4 \pi} \right)^{d/2}  (-m^2)^{-\alpha- \beta}{\Gamma(\alpha+ \beta-d/2) \Gamma(d-\alpha-2 \beta) 
\over\Gamma(\alpha) \Gamma(d-\alpha-\beta)} , \\
\widetilde \CI(\alpha, \beta; m^2, m^2)&=\ri \left( {m^2 \over 4 \pi} \right)^{d/2} (-m^2)^{-\alpha- \beta}{\Gamma(\alpha+ \beta-d/2) \Gamma(d+1-\alpha-2 \beta) 
\over\Gamma(\alpha) \Gamma(d+1-\alpha-\beta)}. 
\ea
\ee
\sectiono{Summing over bubbles}

\label{pmp-trick}

In \cite{pmp, pm}, a useful technique was introduced to obtain the renormalized
perturbative series associated to a chain of bubbles. Here we summarize the main results.

Let us consider a sum of corrections in the bare 't Hooft coupling $\lambda_0$, of order $1/N$:
\be
\sum_{n \ge n_0} a_n (p^2)   \lambda_0^n \ ,
\ee
where $n_0 \ge 1$. The choice of $\lambda_0$ is such that, at leading order in the $1/N$ expansion, the renormalization function
is given by
\be
\label{zl}
Z^{-1}_\lambda= 1+ {\lambda \over \epsilon} + \CO(N^{-1}) \ .
\ee
Then, the renormalized sum is of the form
\be
\label{ren-sum}
\sum_{n \ge n_0} (\nu^2)^{n\epsilon/2} a_n (p^2)  Z_\lambda^n \lambda^n \ .
\ee
Let us assume that one can find a ``structure function" $F(x,y)$, which is analytic in both arguments at $x=0$, $y=0$, and such that one
can write
\begin{equation}
(\nu^2)^{n\epsilon/2} a_n (p^2) = \frac{F(\epsilon,n\epsilon)}{n\epsilon^n} \ .
\label{eq_structure_function}
\end{equation}
We also define
\be
F(\epsilon,y)= \sum_{j \ge 0} F_j(\epsilon) y^j \ .
\ee
Then, one can show that
\be
\label{main-f}
\ba
&\sum_{n \ge n_0} {F(\epsilon, n \epsilon) \over n \epsilon^n} \lambda^n \left(1+{ \lambda \over\epsilon} \right)^{-n}\\
&= F_0(\epsilon)\log\left(1+{\lambda \over \epsilon} \right)-F_0(\epsilon) \sum_{m=1}^{n_0-1}{(-1)^{m-1}  \over m}  \left( {\lambda \over \epsilon} \right)^m   + \sum_{m \ge n_0} (m-1)! F_m (\epsilon) \lambda^m+ \CO(\epsilon) \ .
\ea
\ee
The first two terms in the second line of (\ref{main-f}) lead to divergences as $\epsilon \to 0$. We note the useful formula
\be
\label{F0div}
\left[ F_0(\epsilon) \log\left(1+ {\lambda \over \epsilon} \right) \right]_{\rm div}=  \int_0^\lambda {F_0(-u) \over u+ \epsilon} \rd u \ ,
\ee
and we can write the divergent part in (\ref{main-f}) as
\be
\label{div-part}
  \int_0^\lambda {F_0(-u) \over u+ \epsilon} \rd u - \sum_{m=1}^{n_0-1} {(-1)^{m-1} \over m} \lambda^m \sum_{k=0}^{m-1} F_{0,k} \epsilon^{k-m}\ ,
  \ee
where the coefficients $F_{0,k}$ are defined by
\be
F_0(\epsilon)=  \sum_{k \ge 0} F_{0,k} \epsilon^k.
\ee
The finite part of the second line in (\ref{main-f}) can be written as
\be
\label{finite-part}
-\int_0^\lambda {F_0(-u) -F_{0,0} \over u} \rd u- \sum_{m=1}^{ n_0-1} {(-1)^{m-1} \over m} F_{0,m} \lambda^m+ \sum_{m \ge n_0} (m-1)! F_m (0) \lambda^m \ .
\ee
We note that the last sum has the form of an inverse Borel transform:
\be
\label{fborel}
\sum_{m \ge n_0} (m-1)! F_m (0) \lambda^m= \int_0^\infty \re^{-y/\lambda} \left( \sum_{m \ge n_0} F_m (0) y^{m-1} \right) \rd y =
\int_0^\infty \re^{-y/\lambda}  {\cal B}(y)  \rd y \ ,
\ee
where 
\be
\label{borelF}
{\cal B}(y)= {F(y)- \sum_{m =0}^{n_0-1} F_m (0) y^{m}  \over y}, \qquad F(y)= F(0,y).
\ee

\bibliographystyle{JHEP}

\linespread{0.6}
\bibliography{biblio-qft}

@article{kronfeld,
    author = "Kronfeld, Andreas S.",
    title = "{The Perturbative pole mass in QCD}",
    eprint = "hep-ph/9805215",
    archivePrefix = "arXiv",
    reportNumber = "FERMILAB-PUB-98-139-T",
    doi = "10.1103/PhysRevD.58.051501",
    journal = "Phys. Rev. D",
    volume = "58",
    pages = "051501",
    year = "1998"
}

@article{qcd-propagators,
    author = "Lavelle, Martin and Oleszczuk, Michael",
    title = "{The Operator product expansion of the QCD propagators}",
    reportNumber = "TPR-92-27, MZ-TH-92-33",
    doi = "10.1142/S0217732392003049",
    journal = "Mod. Phys. Lett. A",
    volume = "7",
    pages = "3617--3630",
    year = "1992"
}

@article{beneke-pole,
    author = "Beneke, Martin",
    title = "{Pole mass renormalon and its ramifications}",
    eprint = "2108.04861",
    archivePrefix = "arXiv",
    primaryClass = "hep-ph",
    reportNumber = "TUM-HEP-1358/21",
    doi = "10.1140/epjs/s11734-021-00268-w",
    journal = "Eur. Phys. J. ST",
    volume = "230",
    number = "12-13",
    pages = "2565--2579",
    year = "2021"
}

@article{sss,
    author = "Schubring, Daniel and Sheu, Chao-Hsiang and Shifman, Mikhail",
    title = "{Treating divergent perturbation theory: Lessons from exactly solvable 2D models at large N}",
    eprint = "2107.11017",
    archivePrefix = "arXiv",
    primaryClass = "hep-th",
    doi = "10.1103/PhysRevD.104.085016",
    journal = "Phys. Rev. D",
    volume = "104",
    number = "8",
    pages = "085016",
    year = "2021"
}

@article{lm-nlsm,
    author = "Liu, Yizhuang and Mari{\~n}o, Marcos",
    title = "{Trans-series from condensates in the non-linear sigma model}",
    eprint = "2507.02605",
    archivePrefix = "arXiv",
    primaryClass = "hep-th",
    doi = "10.21468/SciPostPhys.20.6.156",
    journal = "SciPost Phys.",
    volume = "20",
    pages = "156",
    year = "2026"
}

@article{liu-on,
    author = "Liu, Yizhuang",
    title = "{Renormalon-like factorial enhancements to power expansion/OPE in a super-renormalizable 2D O(N) quartic model}",
    eprint = "2502.02031",
    archivePrefix = "arXiv",
    primaryClass = "hep-th",
    doi = "10.1007/JHEP08(2025)014",
    journal = "JHEP",
    volume = "08",
    pages = "014",
    year = "2025"
}

@article{pdr,
	author = {Pascual, Pedro and de Rafael, Eduardo},
	doi = {10.1007/BF01548609},
	journal = {Z. Phys. C},
	pages = {127},
	reportnumber = {CPT-81/P-1316},
	title = {{Gluonic Corrections to Quark Vacuum Condensate Contributions to Two Point Functions in {QCD}}},
	volume = {12},
	year = {1982}}

@article{anatomy,
    author = "Mari\~no, Marcos",
    title = "{Anatomy of the simplest renormalon}",
    eprint = "2504.12044",
    archivePrefix = "arXiv",
    primaryClass = "hep-th",
    doi = "10.21468/SciPostPhys.21.3.056",
    journal = "SciPost Phys.",
    volume = "21",
    pages = "056",
    year = "2026"
}

@article{liu,
    author = "Liu, Yizhuang",
    title = "{Bjorken and threshold asymptotics of a space-like structure function in the 2D U(N) Gross-Neveu model}",
    eprint = "2403.06787",
    archivePrefix = "arXiv",
    primaryClass = "hep-th",
    doi = "10.1007/JHEP09(2024)093",
    journal = "JHEP",
    volume = "09",
    pages = "093",
    year = "2024"
}

@article{parisi-renormalons,
	author = {Parisi, G.},
	doi = {10.1016/0550-3213(79)90298-0},
	journal = {Nucl. Phys. B},
	pages = {163--172},
	reportnumber = {LPTENS 78/11},
	title = {{On Infrared Divergences}},
	volume = {150},
	year = {1979}}

@article{svz,
	author = {Shifman, Mikhail A. and Vainshtein, A. I. and Zakharov, Valentin I.},
	doi = {10.1016/0550-3213(79)90022-1},
	journal = {Nucl. Phys. B},
	pages = {385--447},
	reportnumber = {ITEP-73-1978, ITEP-80-1978},
	title = {{QCD and Resonance Physics. Theoretical Foundations}},
	volume = {147},
	year = {1979}}

@article{shifmanetal,
	archiveprefix = {arXiv},
	author = {Bigi, Ikaros I. Y. and Shifman, Mikhail A. and Uraltsev, N. G. and Vainshtein, A. I.},
	doi = {10.1103/PhysRevD.50.2234},
	eprint = {hep-ph/9402360},
	journal = {Phys. Rev. D},
	pages = {2234--2246},
	reportnumber = {TPI-MINN-94-4-T, UMN-TH-1239-94, CERN-TH-7171-94, UND-HEP-94-BIG03},
	title = {{The Pole mass of the heavy quark. Perturbation theory and beyond}},
	volume = {50},
	year = {1994}}

@article{beneke-braun2,
	archiveprefix = {arXiv},
	author = {Beneke, M. and Braun, Vladimir M.},
	doi = {10.1016/0550-3213(94)90314-X},
	eprint = {hep-ph/9402364},
	journal = {Nucl. Phys. B},
	pages = {301--343},
	reportnumber = {MPI-PHT-94-9, UM-TH-94-4},
	title = {{Heavy quark effective theory beyond perturbation theory: Renormalons, the pole mass and the residual mass term}},
	volume = {426},
	year = {1994}}

@article{tarrach,
	author = {Tarrach, R.},
	doi = {10.1016/0550-3213(81)90140-1},
	journal = {Nucl. Phys. B},
	pages = {384--396},
	reportnumber = {CPT-80/P-1215},
	title = {{The Pole Mass in Perturbative QCD}},
	volume = {183},
	year = {1981}}

@article{elias2,
	author = {Elias, V. and Steele, Thomas G. and Scadron, M. D. and Tarrach, R.},
	doi = {10.1103/PhysRevD.34.3537},
	journal = {Phys. Rev. D},
	pages = {3537},
	reportnumber = {Print-86-0318 (WESTERN ONTARIO)},
	title = {{Truncation of the Operator Product Expansion for the $ < q \bar{q} > $ Condensate Component of the Quark Mass}},
	volume = {34},
	year = {1986}}

@article{elias3,
	author = {Elias, V. and Steele, Thomas G. and Scadron, M. D.},
	doi = {10.1103/PhysRevD.38.1584},
	journal = {Phys. Rev. D},
	pages = {1584},
	reportnumber = {PRINT-87-0467 (WESTERN-ONTARIO)},
	title = {{$q \bar{q}$ and Higher Dimensional Condensate Contributions to the Nonperturbative Quark Mass}},
	volume = {38},
	year = {1988}}

@article{elias4,
	author = {Elias, V.},
	doi = {10.1139/p86-110},
	journal = {Can. J. Phys.},
	pages = {595--601},
	reportnumber = {Print-86-0317 (WESTERN ONTARIO)},
	title = {{Quantum Field Theory and the Dynamical Generation of Quark Masses}},
	volume = {64},
	year = {1986}}

@article{bbk,
	archiveprefix = {arXiv},
	author = {Beneke, M. and Braun, Vladimir M. and Kivel, N.},
	doi = {10.1016/S0370-2693(98)01339-2},
	eprint = {hep-ph/9809287},
	journal = {Phys. Lett.},
	pages = {308-316},
	primaryclass = {hep-ph},
	reportnumber = {CERN-TH-98-290, NORDITA-98-58-HE},
	slaccitation = {%%CITATION = HEP-PH/9809287;%%},
	title = {{The Operator product expansion, nonperturbative couplings and the Landau pole: Lessons from the O(N) sigma model}},
	volume = {B443},
	year = {1998}}

@article{david2,
	author = {David, F.},
	doi = {10.1016/0550-3213(84)90235-9},
	journal = {Nucl. Phys.},
	pages = {237-251},
	reportnumber = {SACLAY-SPH-T-83-059},
	slaccitation = {%%CITATION = NUPHA,B234,237;%%},
	title = {{On the Ambiguity of Composite Operators, IR Renormalons and the Status of the Operator Product Expansion}},
	volume = {B234},
	year = {1984}}

@article{beneke,
	archiveprefix = {arXiv},
	author = {Beneke, M.},
	doi = {10.1016/S0370-1573(98)00130-6},
	eprint = {hep-ph/9807443},
	journal = {Phys. Rept.},
	pages = {1-142},
	primaryclass = {hep-ph},
	reportnumber = {CERN-TH-98-233},
	slaccitation = {%%CITATION = HEP-PH/9807443;%%},
	title = {{Renormalons}},
	volume = {317},
	year = {1999}}

@article{est,
	author = {Elias, V. and Scadron, M. D. and Tarrach, R.},
	doi = {10.1016/0370-2693(85)91082-2},
	journal = {Phys. Lett. B},
	pages = {176--180},
	title = {{G}AUGE INDEPENDENCE OF THE OPERATOR PRODUCT NONPERTURBATIVE QUARK MASS},
	volume = {162},
	year = {1985}}

@article{politzer,
	author = {Politzer, H. David},
	doi = {10.1016/0550-3213(76)90405-3},
	journal = {Nucl. Phys. B},
	pages = {397--406},
	reportnumber = {HUTP-76/A159},
	title = {{Effective Quark Masses in the Chiral Limit}},
	volume = {117},
	year = {1976}}

@book{pascual-tarrach,
	author = {Pascual, P. and Tarrach, R.},
	publisher = {Springer-Verlag},
	series = {Lecture Notes in Physics},
	title = {{QCD: renormalization for the practicioner}},
	volume = {194},
	year = {1984}}

@article{cr-review,
	author = {Campostrini, Massimo and Rossi, Paolo},
	doi = {10.1007/BF02730034},
	journal = {Riv. Nuovo Cim.},
	number = {6},
	pages = {1--111},
	title = {{The 1/N expansion of two-dimensional spin models}},
	volume = {16},
	year = {1993}}

@article{cr-dr,
	author = {Campostrini, Massimo and Rossi, Paolo},
	doi = {10.1142/S0217751X92001459},
	journal = {Int. J. Mod. Phys. A},
	pages = {3265--3290},
	reportnumber = {IFUP-TH-9-91},
	title = {{Dimensional regularization in the 1/N expansion}},
	volume = {7},
	year = {1992}}

@article{pmp,
	author = {Palanques-Mestre, A. and Pascual, P.},
	doi = {10.1007/BF01212398},
	journal = {Commun. Math. Phys.},
	pages = {277},
	reportnumber = {Print-83-0417 (BARCELONA)},
	title = {{The 1/$N_f$ Expansion of the $\gamma$ and beta Functions in QED}},
	volume = {95},
	year = {1984}}

@article{pm,
	author = {Palanques-Mestre, A.},
	doi = {10.1007/BF01552504},
	journal = {Z. Phys. C},
	pages = {255},
	reportnumber = {Print-86-0264 (BARCELONA)},
	title = {{Renormalons in QED}},
	volume = {32},
	year = {1986}}

@article{gross-neveu,
	author = {Gross, David J. and Neveu, Andre},
	doi = {10.1103/PhysRevD.10.3235},
	journal = {Phys. Rev.},
	pages = {3235},
	reportnumber = {COO-2220-19},
	slaccitation = {%%CITATION = PHRVA,D10,3235;%%},
	title = {{Dynamical Symmetry Breaking in Asymptotically Free Field Theories}},
	volume = {D10},
	year = {1974}}

@article{fnw1,
	author = {Forgacs, P. and Niedermayer, F. and Weisz, P.},
	doi = {10.1016/0550-3213(91)90044-X},
	journal = {Nucl. Phys.},
	pages = {123-143},
	reportnumber = {MPI-PH-91-37},
	slaccitation = {%%CITATION = NUPHA,B367,123;%%},
	title = {{The Exact mass gap of the Gross-Neveu model. 1. The Thermodynamic Bethe ansatz}},
	volume = {B367},
	year = {1991}}

@article{ft,
	archiveprefix = {arXiv},
	author = {Faddeev, Ludwig D. and Takhtajan, Leon A.},
	eprint = {1408.0307},
	primaryclass = {math.SP},
	slaccitation = {%%CITATION = ARXIV:1408.0307;%%},
	title = {{On the spectral theory of one functional-difference operator from conformal field theory}},
	year = {2014}}

@article{n,
	archiveprefix = {arXiv},
	author = {Nekrasov, Nikita A.},
	doi = {10.4310/ATMP.2003.v7.n5.a4},
	eprint = {hep-th/0206161},
	journal = {Adv. Theor. Math. Phys.},
	pages = {831-864},
	primaryclass = {hep-th},
	reportnumber = {ITEP-TH-22-02, IHES-P-04-22},
	slaccitation = {%%CITATION = HEP-TH/0206161;%%},
	title = {{Seiberg--Witten prepotential from instanton counting}},
	volume = {7},
	year = {2004}}

@article{mm-condensates,
	archiveprefix = {arXiv},
	author = {Mari\~no, Marcos and Miravitllas, Ramon},
	doi = {10.21468/SciPostPhys.18.3.101},
	eprint = {2402.19356},
	journal = {SciPost Phys.},
	number = {3},
	pages = {101},
	primaryclass = {hep-th},
	title = {{Trans-series from condensates}},
	volume = {18},
	year = {2025}}

\end{document}